\documentclass[superscriptaddress,aps,prd,twocolumn,showpacs,nofootinbib]{revtex4-1}
\usepackage{etex}
\usepackage{amsmath,amssymb,amsthm}
\usepackage{ragged2e}
\usepackage[colorlinks=true,citecolor=blue,urlcolor=blue]{hyperref}
\usepackage[pdftex]{graphicx}  % need for figures
\usepackage{verbatim}   % useful for program listings
\usepackage{color}      % use if color is used in text
\usepackage{subfigure}  % use for side-by-side figures
\usepackage{hyperref}   % use for hypertext links, including those to external documents and URLs
\usepackage{mathrsfs}
\usepackage{booktabs}
\usepackage[font=footnotesize,labelfont=bf]{caption} % Required for specifying captions to tables and figures
\usepackage{wrapfig}
\begin{document}
\title{ Perturbative correction terms to electromagnetic self-force due to metric perturbation : astrophysical and cosmological implications }
\author{Arnab Sarkar}         \email{arnabsarkar@bose.res.in, arnab.sarkar14@gmail.com}
\affiliation{Department of Astrophysics and Cosmology, 
S. N. Bose National Centre for Basic Sciences, JD Block, Sector III, 
Salt lake city, Kolkata-700106, India}
\author{Amna Ali }  \email{amnaalig@gmail.com}
\affiliation{Department of Mathematics, Jadavpur University, Kolkata-700032, India}
\author{Salah Nasri}   \email{snasri@uaeu.ac.ae}
\affiliation{Department of Physics, United Arab Emirates University, P.0.BOX 15551, Al- Ain, United Arab Emirates}
\affiliation{International Centre for Theoretical Physics, Strada Costiera 11, I-34151 Trieste, Italy}
\date{\today}
\begin{abstract}
\begin{center}
\textbf{Abstract}
\end{center}
\begin{small}
We consider the equation of motion of a charged particle or a charged compact object in curved space-time, under the reaction of electromagnetic radiation and also consider a physical situation such that the charged particle or compact object emits gravitational radiation, thereby gravitational radiation reaction also acts on it. We investigate the effect of this metric perturbation i.e. the gravitational radiation on the electromagnetic self-force.   
 We show that, besides the interaction terms derived by P. Zimmerman and E. Poisson \cite{Zimmerman_et_al}, additional perturbative terms are generated, which are linear in metric perturbation and are generated due to perturbation of the electromagnetic self-force by the metric perturbation. We discuss the conditions of significance of these perturbative terms and also the interaction terms with respect to the gravitational self-force in various astrophysical and cosmological cases ; such as the motion of charged particles around black holes, some extreme mass-ratio inspirals (EMRIs) involving sufficiently accelerated motion of charged stars (specially neutron stars) or charged stellar mass black holes around supermassive black holes, and motion of charged particles around charged primordial black holes formed in the early Universe etc.. We find that in some astrophysical and cosmological cases these perturbative terms can have significant effect in comparison with the gravitational radiation-reaction term.
%We plan to discuss the effect of interaction of gravitational radiation-reaction and electromagnetic 'Tail term' in another work later. 
%\keywords{Neutron Stars, Strange matter, Modified Gravity}%Use showkeys class option if keyword
%display desired
\end{small}
\end{abstract}
\maketitle
%______________________________________________________________________________________________________
%______________________________________________________________________________________________________
\begin{small}
\section{Introduction}
The motion of a point charge in flat space-time was one of the main topics of research in physics from as early as 1930s. Many pioneering physicists like Lorentz, Abrahams, Poincare and Dirac contributed to the development of the subject from its early onset \cite{Dirac}. %In 1960,
DeWitt and Brehme generalized Dirac's result to curved spacetimes, and, for the first time, gave a precise derivation of Electromagnetic self-force \cite{DeWittBrehme}. Later, Hobbs applied vierbein treatment to derive their equations and found that their results must be corrected by a term involving the Ricci tensor \cite{Hobbs}. % Later while studying the motion of point %masses in curved spacetime, the 
The rigorous derivation of the electromagnetic self-force was given by Samuel E. Gralla et al, in their work in 2009 \cite{Gralla_et_al}. \\
Similar counterpart of electromagnetic self-force in gravity viz. the `Gravitational self-force' also was derived, first by Mino, Sasaki, and Tanaka \cite{MiSaTa}, and then by Quinn and Wald using a different method \cite{QuWa}. 
%Thereafter various developments has been taking place in this subject till now. 
\\
In this work, we consider an interesting case where both the electromagnetic and gravitational self-forces are present ; and both of them produce corresponding radiation reactions in the motion. We consider the equation of motion of a charged particle in curved space-time under electromagnetic radiaton reaction, with external Lorentz force too and consider a physical situation such that the particle emits gravitational radiation, which perturbs the surrounding space-time. We then follow the procedure of deriving the MiSaTaQuWa equation for this equation of motion.\\ 
Here the term `particle' does not strictly mean that it has very tiny size like elementary particles ; the mass should be centralized enough so that the equations of motion of point-particles can be applied. In this sense, a compact object like a neutron star or a stellar mass black hole orbiting a supermassive black hole can also be treated by a point-particle equation of motion. Although there are problems with point-particle notion when terms with second order metric perturbations are considered in calculation, we will not be facing it as we are considering linear metric perturbations only. \\
In this context the work by Peter Zimmerman and Eric Poisson \cite{Zimmerman_et_al} is very important, as it is probably the first work treating the case where electromagntic and gravitational self-forces act together, while deriving their interaction terms. The authors in this work have considered not only the electrovac space-time, but also a more general case of scalarvac space-time, where the metric of the background space-time is a solution of the Einstein's equation in the presence a scalar field.  
We show that in comparison to the case where only gravitational self-force is present, the presence of electromagnetic self-force with it, adds not only the interaction terms but also several extra perturbative terms in the equation of motion.
 %which are generated due to perturbation of the electromagnetic self-force by the metric perturbations.
If we see the order only in terms of the charge and mass of the charged particle, then the order of these terms are proportional to $ q^{2}m$ for force and to $q^{2} $ for acceleration ; $q $ being the electric charge of the particle and $m  $ being its mass. We shall discuss the issue of orders of terms in detail, in section \ref{Sec3} in this work. 
%If they were the interaction terms of elctromagnetic and gravitational self-forces, then they would have to be proportional to the $qm $, where $m$ is the mass of the particle.
 We interpret that these additional terms are just the perturbations to the electromagnetic self-force, due to the gravitational radiation or metric perturbations emitted by the system. We have considered these perturbative terms to the first order of the metric perturbations. \\
% just like the gravitational self-force. \\
 We investigate some astrophysical systems and cosmological cases, where these additional perturbative terms produced from the electromagnetic self-force, are significant in comparison with the gravitational self-force. In this case it is to be noted that although the gravitational self-force and these perturbative terms have a distinct difference to the sense that the first one is a purely gravitational aspect, while the origin of the perturbative terms are electromagnetic ; yet, they have the similarity that they contain the metric perturbations (to the linear order in this case) as they are generated due to metric perturbations. The motive of this comparison is that this will help to identify the cases where the perturbative terms generated from electromagnetic self-force would have significance or would dominate over the gravitational self-force and vice-versa. The equations of motion in the corresponding cases can be simplified or approximated accordingly.\\
We find that although in a few cases it is possible that these perturbative terms can be significant in comparison with the gravitational self-force, the special interest comes out to be in the cases of charged particle's motions around primordial black holes within certain mass range, which were produced in early Universe by direct gravitational collapse of sufficiently deep density perturbations.\\
We have discussed the orthogonality, with the four-velocity of the particle, of different terms present in the radiation reaction, in Appendix-1. There are other two Appendices. In Appendix-2, we have explained why we have neglected the perturbations originated from the term containing Ricci-tensors and in Appendix-3 we have discussed certain issues related to the comparison of different parts within the perturbative correction terms, with the gravitational self-force term.
 \\
We have not discussed the additional perturbative terms generated from the electromagnetic `Tail term' due to the metric perturbations, in this work, which will be left for a separate future work.
% We hope to do that later in a different work with a detailed analysis.   
 In this work, we have preliminarily written the electromagnetic self-forces in Gaussian units with c (speed of light in vacuum ) = 1 ; but while estimating some numerical quantities related to it, we have converted this to S.I. units system.   
%________________________________________________________________________________________________________________
\section{Overcoming singularity of the retarded metric perturbation on the world line of the particle and the gauge fixing :}
One main issue of our work is that the physical or retarded metric perturbation $h_{\mu \nu}^{ret} $ emitted from the charged particle is singular at the particle, or in other words, it is singular on the world-line of the particle. As our work deals with the world-line of the particle, we must address this singularity of the metric perturbation emitted form the particle.\\
For overcoming the problem of singularity of the retarded metric perturbation on the world-line of the particle, we follow the \textit{Detweiler-Whiting formalism} \cite{Detweiler-Whiting}, in which Detweiler and Whiting proposed a reformulation of the perturbed motion, where instead of breaking the overall retarded perturbation into `Direct' and `Tail' parts, they decomposed it into `Singular' and `Regular' parts :   
\begin{equation} \label{r1}
h_{\mu \nu}^{ret} = h_{\mu \nu}^{S} + h_{\mu \nu}^{R} \, , 
\end{equation} 
where the singular part $h_{\mu \nu}^{S}$ is responsible for the singular behaviour of the retarded perturbation on the world-line, while it does not generate any self-force or does not affect the motion of the particle. On the otherhand, the regular part $ h_{\mu \nu}^{R} $ is a smooth solution of the perturbation equations and this exerts the identical self-force, as generated by the overall retarded perturbation.
The \textit{Detweiler-Whiting formalism} indicates the fact that the particle effectively moves along a geodesic\footnote{it is to be noted that we can use the term geodesic only in case of absence of external forces ;} of a smooth perturbed space-time with the metric $g'_{\mu \nu} = g_{\mu \nu} + h_{\mu \nu}^{R} $\cite{Detweiler-Whiting, Barack, Poisson}, where $ g_{\mu \nu}$ is the unperturbed metric.\\ 
On the world-line of the particle, the regular part of the perturbation satisfies \cite{Poisson} 
\begin{equation} \label{r2}
\nabla_{\lambda} h_{\mu \nu}^{R} = -4m(u_{(\mu} R_{\nu ) \rho  \lambda \eta} + R_{\mu \rho \nu \eta} u_{\lambda} )u^{\rho} u^{\eta} + \bar{h}_{\mu \nu ; \lambda}^{Tail}   \, , 
\end{equation} 
where the $ \bar{h}_{\mu \nu ; \lambda}^{Tail} $ is given by (in trace-reversed form) : 
\begin{equation} \label{r3}
\begin{aligned}
\bar{h}_{\mu \nu ; \lambda}^{Tail}  = 4m \int_{- \infty}^{\tau^{-}}  \nabla_{\lambda} \Big(G_{+ \mu \nu \mu'' \nu''} -
 \frac{1}{2}  g_{\mu \nu } G_{+  \rho \mu'' \nu''}^{\, \rho}   \Big)   (z(\tau), z(\tau''))
 \\
  u''^{\mu} u''^{\nu} d\tau'' \, . 
\end{aligned}
\end{equation}    
Therefore, we shall only work with the regular metric perturbation $ h_{\mu \nu}^{R} $, thereby eliminating the singular behaviour of the retarded metric perturbation, emitted from the particle, on its world-line and all the consequent general relativistic perturbation quantities will be in terms of $ h_{\mu \nu}^{R}$ .\\
In this work, it is pertinent to be mentioned that, although in case of electromagnetic self-force, the point particle approximation is valid, but in case of general relativity, the point particle concept fails at non-linear orders. But, as the point particle concept has no problem with the linear orders of metric perturbation, which we are considering here, we can stick to this concept. However, it is worth noting that if any calculation is necessary in the non-linear orders of metric perturbation, where instead of point-particle notion the object of interest is a compact one, one has to employ a different method, known as `Puncture method' \cite{Barack-Golbourn, Barack-Golbourn-Sago-2,  Vega-Detweiler}. In this method the retarded metric perturbation is divided into two parts known as `Puncture part' $ h_{\mu \nu}^{\mathcal{P}} $ and `Residual part' $ h_{\mu \nu}^{\mathcal{R}} $. For detailed discussion on this method, previous references \cite{Barack-Golbourn, Barack-Golbourn-Sago-2, Vega-Detweiler} and the review article by L. Barack and A. Pound \cite{Barack-Pound} can be consulted. We donot go into detail about this `Puncture method' here, as it is not required for our analysis done with first-order metric perturbations. \\
Another issue in our work is that as the gravitational self-force and the metric perturbation are both gauge-dependent quantities, we have to fix the gauge to describe this physical effect meaningfully. We here choose it to be the \textit{Lorenz-gauge}, in which many preliminary and foundational results in gravitational self-force had been obtained. In terms of the trace-reversed form of the metric perturbation : 
\begin{equation} \label{r4}
\bar{h}_{\mu \nu} = h^{ret}_{\mu \nu} - \frac{1}{2} g_{\mu \nu } h^{ret} 
\end{equation} 
($h^{ret} = g^{\alpha \beta } h^{ret}_{\alpha \beta} $), the Lorenz-gauge condition is given by :
\begin{equation} \label{r5}
\nabla^{\mu} \bar{h}_{\mu \nu } = 0  \, .
\end{equation}
Here, a confusion may arise that we will be working with the regular part of the perturbation $h^{R}_{\mu \nu} $, while the gauge-condition is in terms of the trace-reversed version of the overall perturbation $h^{ret}_{\mu \nu}$. But, there is no problem with this issue, as what happens is actually that the singular part $h^{S}_{\mu\nu } $ remains invariant under a smooth gauge-transformation, while only the regular part $h^{R}_{\mu \nu} $ changes.   \\
Furthermore, in our case, as there is electromagnetic radiation emitted from the charged particle in the space-time surrounding it and in the most general case there is also the external electromagnetic field, hence the metric actually satisfies the Einstein-Maxwell equation, not the vacuum Einstein equation. If we consider an EMRI, 
%and take the origin of the coordinate system at the centre-of-mass of the larger massive body around which the much smaller massive body is revolving, 
where the smaller component is charged, 
then the metric outside the larger massive body in the EMRI satisfies the Einstein's equation :
\begin{equation} \label{r6}
R_{\mu \nu} - \frac{1}{2} g_{\mu \nu} R = 8 \pi G T_{\mu \nu } \, , 
\end{equation}  
where the energy-momentum tensor $ T_{\mu \nu }  $ of the source outside the larger massive body contains two components : the energy-momentum tensor of the smaller charged massive body itself and the energy momentum tensor of the electromagnetic radiation emitted from the smaller charged massive body (if there be any external electromagnetic field then its energy-momentum tensor is also to be added). With the point-particle approximation, the energy momentum tensor of the smaller massive body can be represented by a Dirac-Delta function.\footnote{it is to be noted that this point-particle approximation of the smaller massive body and hence the corresponding Dirac-Delta function representation of its energy-momentum tensor would not be valid when second order metric perturbations would be considered, as we have already stated.} Hence, outside both the larger and smaller massive bodies, the only energy-momentum is of the electromagnetic radiation, or in other words, the metric there satisfies the Einstein-Maxwell equation :
\begin{equation} \label{r7}
R_{\mu \nu} - \frac{1}{2} g_{\mu \nu} R = \frac{8 \pi G}{ \mu_{o}}  (F_{\mu}^{\,\, \eta } F_{\nu\eta} - \frac{1}{4} g_{\mu \nu} F^{2} ) \, ,
\end{equation}
where $\mu_{0} $ is the permeability in vacuum and $ F_{\mu \nu}$ is the electromagnetic field-strength tensor. So, the metric perturbations outside the larger massive body would satisfy the first order perturbation-equation of the background equation \ref{r7} . 
%But, for simplicity here we assume that the electromagnetic field-strength tensor, which is acting as the source-term in the Einstein-Maxwell's equation, is negligible or in other way, that is too feeble such that the equation \ref{r7} can be taken effectively as Einstein's equation in vacuum. 
%________________________________________________________________________________________________________________
\section{Order of the perturbative terms:} \label{Sec3}
Samuel E. Gralla et al gave a rigorous derivation of the electromagnetic self-force in their work in 2009 \cite{Gralla_et_al}. In their work, they followed an approach called ``asymptotic self-similar manner". In this approach, about the worldline of the charged body as $ \lambda \rightarrow 0 $ ($\lambda $ is a small parameter measuring the size of the charge and mass, not a parameter along the worldline \cite{Gralla_et_al}), the charge $q $ and mass $m $ of the charged body (of whose equation of motion is to be studied) tends to zero, but the charge-to-mass ratio $q/m $ tends to a well-defined limit. This approach is considered mainly to tackle the difficulty with the point particle description of the charged body, when the limit is taken to zero-size in a straightforward way and also to avoid the problems associated with the body's finite-size consideration. As a result of following this asymptotic self-similar approach, in most of the cases the scaling of $q $ and $m $ are treated on equal footing i.e. it is presumed that $ q \sim m $, which may not be the case in some astrophysical and cosmological scenarios.\\ 
Furthermore, when the order of a certain term in the equation of motion of the charged object is spoken about, it is often estimated and compared with the other terms only in terms of $q $ and $m $. But in various cases, any other physical quantity present in that certain term, e.g. the four velocity, four acceleration and rate of change of four acceleration etc., may have such a huge order that they must have to be considered. Otherwise the estimation or comparison of the orders of these terms would turn out to be just incorrect. \\
In our work, we discuss some astrophysical and cosmological cases where the orders of these quantities like four velocity, four acceleration or rate of change of four acceleration are so huge that the overall orders of the concerned terms can not be judged only in terms of $q $ and $m $. In fact, in this work we study all the perturbative terms which are of linear order in metric perturbation $h^{R}_{\alpha \beta} $, instead of designating the order of perturbation both in terms of the electromagnetic perturbation and gravitational or metric perturbation.
% Generally, the electromagnetic field produced by the charged body is seen as perturbation in the background electromagnetic field.     
\\
Generally the electromagntic self-force is seen as a perturbation over the external lorentz force and hence it may seem that the electromagntic self-force can not be compared with the gravitational force viz. the main newtonian-part of the gravitational force. But, quite recently A. Tursunov et al has shown \cite{Tursunov_et_al} that for a charged particle with charge q and with relativistic speed, the electromagnetic self-force, which is of the order of $\sim q^{4}B^{2}/m^{2} $, can have same order of magnitude as that of Newtonian gravitational force (of the order $ \sim GMm/r^{2} $), when the charged particle is moving around a supermassive black hole of mass $\sim 10^{9}M_{\odot} $ ( $  M_{\odot}$ is the usual symbol of Solar-mass), in presence of a magnetic field of $B \sim 10^{4} $ G. \footnote{According to the works \cite{Piotrovich_et_al} and \cite{Baczko_et_al} The characteristic values of magnetic field near supermassive black holes of mass $\sim 10^{9}M_{\odot} $ is B $\sim 10^{4} $ G .} So, as it is possible in a practical case that the electromagnetic self-force can be of same order of magnitude with the newtonian gravitational force, it may also be possible for the perturbations in the eletromagntic self-force caused by metric perturbations to have similar orders of magnitude as that of the gravitational self-force.\\
At last it is of utter importance to remember the fact that we have considered here metric perturbations originated due to the motion of the charged body itself. If there is external metric perturbation, when we speak about the orders in terms of charge and mass only, then the perturbative terms of the electromagnetic self-force originated from that external metric fluctuations would not have the order $q^{2}m $, instead they would have the order $q^{2} $, same as that of the electromagnetic self-force (As the $ h_{\alpha \beta} $ then would not be of the order of $m$). So, in that case these terms would have similar order with the gravitational self-force, which is of the order of $m^{2} $. Although in that case, for determination of the correct order would require the knowledge of the mass of that external souce of the gravitational radiation. In the case of external gravitational wave, we shall not need the \textit{Detweiler-Whiting reformulation} to break the metric perturbation into singular and regular parts, as the external metric perturbation will not be singular on the world-line of the particle. However, in this work we shall stick to the case where the gravitational radiation is generated from the charged particle itself.          
%________________________________________________________________________________________________________________  
\section{ The equation of motion of a charged particle in curved space-time under electromagnetic radiation reaction and applying metric perturbation to it : }
Here we consider the explicit form of the equation of motion of a charged particle or compact object in curved space-time under the reaction of the electromagnetic radiation emitted by itself\cite{Poisson, Tursunov_et_al} and also consider the reaction of gravitational radiation, generated due to the motion of the particle around a comparatively very bigger massive compact object, preferably a black hole. The gravitational radiation emitted from the system creates a perturbation of space-time and that would have an effect on the motion of the particle, which is the reason behind the gravitational radiation reaction. Let $\tau' $ be the proper time associated with this perturbed metric of the particle, and $ \tau $ the proper time for the unperturbed metric, without the reaction of gravitational radiation.
 We denote the unperturbed and perturbed metric as $g_{\mu \nu} $ and $g'_{\mu \nu} $ respectively, and physical or `retarded' part of the metric-perturbation (without the `advanced' part of it), i.e. here the gravitational radiation emitted from the system, as $h_{\mu \nu}^{ret} $. To tackle the singularity of the metric perturbation on the world-line of the particle, we take the perturbed metric on the world-line as the effective metric : $g'_{\mu \nu} = g_{\mu \nu} + h_{\mu \nu}^{R}  $, as we have already explained in the previous section 2. 
%The  covariant derivation of the particle's 4-velocity vector on the left  expression of the covariant derivative of $u^{\mu}$ with respect to the proper time geodesic equation in the perturbed metric is given by :
The equation of motion of the charged particle in the perturbed metric is given by :  
\begin{equation} \label{1.1}
\begin{aligned}
\frac{D u'^{\mu}}{d \tau' } = \frac{q}{m} F'^{\mu}_{\, \nu}u'^{\nu} + \frac{2q^{2}}{3m} \Big( \frac{D^{2} u'^{\mu}}{d\tau'^{2}} + u'^{\mu} u'_{\nu} \frac{D^{2} u'^{\nu}}{d \tau'^{2}} \Big)   + 
\\  
\frac{q^{2}}{3m} (R'^{\mu}_{\lambda} u'^{\lambda} + R'^{\nu}_{\lambda} u'^{\lambda} u'^{\mu} u'_{\nu} ) +    \frac{2q^{2}}{m} f'^{\mu \nu}_{Tail} u'_{\nu}  .
\end{aligned}
\end{equation} 
Here, $\frac{D u'^{\mu}}{d \tau'}$ is the covariant derivative of the particle's 4-velocity with respect to $\tau'$, given by
\begin{equation} \label{1.2.1}
\frac{D u'^{\mu}}{d \tau' } =\frac{D}{d\tau'} \frac{d x^{\mu}}{d \tau'} = \frac{d^{2} x^{\mu}}{d \tau'^{2}} + \Gamma'^{\mu}_{\nu \rho} \frac{d x^{\nu}}{d \tau'} \frac{d x^{\rho}}{d \tau'} .
\end{equation}
On the right-hand side  of  Eqn.(\ref{1.1}), the first term  is the Lorentz force acting on the particle, the second term is the electromagnetic radiation reaction in curved space time, the third term is due to the interaction of the particle with the surrounding matter (if there be any) and the fourth term is the `tail term' of the electromagnetic radiation in curved space-time. 
%The  expression of  $\frac{D u'^{\mu}}{d \tau' } $ is given by :
%\begin{equation} \label{1.2.1}
%\frac{D u'^{\mu}}{d \tau' } = \frac{D^{2} x^{\mu}}{d \tau'^{2}} = \frac{d^{2} x^{\mu}}{d \tau'^{2}} + \Gamma'^{\mu}_{\nu \rho} \frac{d x^{\nu}}{d \tau'} \frac{d x^{\rho}}{d \tau'} 
%\end{equation}
%In the work by Tursunov et al \cite{Tursunov_et_al}, the authors have given the expanded expression of
The quantity $\frac{D^{2} u^{\mu} }{d \tau^{2}} $ can be expanded  as \cite{Tursunov_et_al}:
\begin{equation} \label{1.2.2}
\begin{aligned}
\frac{D^{2} u^{\mu} }{d \tau^{2}} = \frac{d^{2} u^{\mu}}{d \tau^{2}} + \partial_{\gamma} \Gamma^{\mu}_{\alpha \beta } u^{\alpha} u^{\beta} u^{\gamma} + 
\\
 3 \Gamma^{\mu}_{\alpha \beta} u^{\alpha} \frac{d u^{\beta}}{d \tau} + \Gamma^{\mu}_{\alpha \beta } \Gamma^{\beta}_{\rho \sigma} u^{\rho} u^{\sigma} u^{\alpha}     .
\end{aligned}
\end{equation}
\vspace{1cm}\\
Now, we designate the contribution of the metric perturbation by an additive vector $ a^{\mu}$ \footnote{It is to be noted very carefully that it can not be called solely a gravitational radiation reaction term. The reason for this will be clear at last, when we shall get its expression.} in the equation of motion of the particle in unperturbed metric, in the following way (for the method used here, Ref.\cite{Barack} may be consulted): 
\begin{equation} \label{1.3}
\begin{aligned}
\frac{D u^{\mu}}{d \tau} =  \frac{q}{m} F^{\mu}_{\,\, \nu}u^{\nu} + \frac{2q^{2}}{3m} \Big( \frac{D^{2} u^{\mu}}{d\tau^{2}} + u^{\mu} u_{\nu} \frac{D^{2} u^{\nu}}{d \tau^{2}} \Big)   +   
\\
\frac{q^{2}}{3m} (R^{\mu}_{\lambda} u^{\lambda} + R^{\nu}_{\lambda} u^{\lambda} u^{\mu} u_{\nu} ) +    \frac{2q^{2}}{m} f^{\mu \nu}_{Tail} u_{\nu} + a^{\mu}     .
\end{aligned}
\end{equation}
%\vspace{1cm}
%For implementing the method used by Barack in his work \cite{Barack}, we use the following differential relations between the $ \tau $ and $ \tau' $ :
%\begin{center}
%${\displaystyle \frac{d}{d\tau} \equiv \frac{d \tau'}{d \tau} \frac{d}{d \tau'} }$  ; \\
%${\displaystyle \frac{d^{2}}{d\tau^{2}} \equiv \frac{d^{2} \tau'}{d \tau^{2}} \frac{d}{d \tau'} + \Big( \frac{d \tau'}{d \tau} \Big)^{2} \frac{ d^{2}}{d \tau'^{2}} }$   ; \\
%${\displaystyle \frac{d^{3}}{d\tau^{3}} \equiv \frac{d^{3}\tau'}{d \tau^{3}} \frac{d \tau'}{d \tau} \frac{d}{d\tau'} + 3\frac{d \tau'}{d \tau} \frac{d^{2} \tau'}{d \tau^{2}} \frac{d^{2}}{d \tau'^{2}}  + \Big(  \frac{d \tau'}{d %\tau}  \Big)^{3} \frac{d^{3}}{d \tau'^{3}}   }$
%\end{center}
%Writing the expanded expressions of the quantities $\frac{D u^{\mu}}{d \tau}$ and $ \frac{D^{2} u^{\mu}}{d \tau^{2}} $ in the equation \ref{1.3}, we write it in a detailed way for clarity : 
%\begin{equation} \label{1.4.1}
%\begin{aligned}
%\frac{d^{2} x^{\mu}}{d \tau^{2}} + \Gamma^{\mu}_{\nu \rho} \frac{d x^{\nu}}{d \tau} \frac{d x^{\rho}}{d \tau}  = 
% \frac{q}{m} F^{\mu}_{\,\, \nu}u^{\nu} +  \\
% \frac{2 q^{2}}{3 m } \Big( g^{\mu}_{\eta} + u^{\mu}u_{\eta} \Big) \frac{D^{2}u^{\eta}}{d \tau^{2}} 
% +  \frac{q^{2}}{3m} (R^{\mu}_{\lambda} u^{\lambda} + R^{\nu}_{\lambda} u^{\lambda} u^{\mu} u_{\nu} ) + 
% \\
 %  \frac{2q^{2}}{m} f^{\mu \nu}_{Tail} u_{\nu} + a^{\mu}
% \end{aligned}
%\end{equation}
Further substituting the expression of the first and second order covariant derivatives of the four-velocity of the particle, into the above Eqn.(\ref{1.3}), we obtain  % $ \frac{D^{2}u^{\eta}}{d \tau^{2}} $ in the RHS of the above equation we obtain :
\begin{equation} \label{1.4.2}
\begin{aligned}
\frac{d^{2} x^{\mu}}{d \tau^{2}} + \Gamma^{\mu}_{\nu \rho} \frac{d x^{\nu}}{d \tau} \frac{d x^{\rho}}{d \tau}  = 
 \frac{q}{m} F^{\mu}_{\,\, \nu}u^{\nu} +  
 \\
 \frac{2 q^{2}}{3 m } \Big( g^{\mu}_{\eta} + u^{\mu}u_{\eta} \Big) \Big( \frac{d^{2} u^{\eta}}{d \tau^{2}} + \partial_{\gamma} \Gamma^{\eta}_{\alpha \beta } u^{\alpha} u^{\beta} u^{\gamma} + 
 \\
 3 \Gamma^{\eta}_{\alpha \beta} u^{\alpha} \frac{d u^{\beta}}{d \tau} + \Gamma^{\eta}_{\alpha \beta } \Gamma^{\beta}_{\rho \sigma} u^{\rho} u^{\sigma} u^{\alpha}  \Big) +
 \\
   \frac{q^{2}}{3m} (R^{\mu}_{\lambda} u^{\lambda} + R^{\nu}_{\lambda} u^{\lambda} u^{\mu} u_{\nu} ) + 
   \frac{2q^{2}}{m} f^{\mu \nu}_{Tail} u_{\nu} + a^{\mu}    .
\end{aligned}
\end{equation}
Next, we substitute the operators $\frac{d}{d\tau} $, $\frac{d^{2}}{d\tau^{2}}$ and $\frac{d^{3}}{d\tau^{3}} $ in the above Eqn.(\ref{1.4.2}) with the similar ones with respect to $\tau' $ ; and obtain :
\vspace{1cm}\\
\begin{widetext}
\begin{equation}  \label{1.5}
\begin{aligned}
\frac{d^{2} \tau'}{d \tau^{2}} \frac{d x^{\mu}}{d \tau'} + \Big( \frac{d \tau'}{d \tau} \Big)^{2} \frac{ d^{2} x^{\mu}}{d \tau'^{2}} +   \Gamma^{\mu}_{\nu \rho} \Big(\frac{d \tau'}{d \tau} \Big)^{2} \frac{d x^{\nu}}{d \tau} \frac{d x^{\rho}}{d \tau} =   
\frac{q}{m}  F^{\mu}_{\,\, \nu} \frac{d \tau'}{d \tau} \frac{d x^{\nu}}{ d \tau'} + \frac{2 q^{2}}{3 m } \Big( g^{\mu}_{\eta} +  \Big(\frac{d \tau'}{d \tau} \Big)^{2}  \frac{d x^{\mu}}{ d \tau'}  \frac{d x_{\eta}}{ d \tau'} \Big)  
\\
\Big[   \left\lbrace \Big( \frac{d^{3}\tau'}{d \tau^{3}} \frac{d \tau'}{d \tau} \Big)  \frac{d x^{\eta}}{d\tau'} + 3\Big( \frac{d \tau'}{d \tau} \frac{d^{2} \tau'}{d \tau^{2}} \Big)  \frac{d^{2} x^{\eta}}{d \tau'^{2}} +  \Big(  \frac{d \tau'}{d \tau}  \Big)^{3} \frac{d^{3} x^{\eta}}{d \tau'^{3}}  \right\rbrace +  
 \Big(  \frac{d \tau'}{d \tau}  \Big)^{3} \partial_{\gamma} \Gamma^{\eta}_{\alpha \beta}   \frac{d x^{\alpha}}{ d \tau'} \frac{d x^{\beta}}{ d \tau'} \frac{d x^{\gamma}}{ d \tau'}  
 \\
 + 3 \Gamma^{\eta}_{\alpha \beta}  \frac{d \tau'}{d \tau}   \frac{d x^{\alpha}}{d \tau'} \left\lbrace \frac{d^{2} \tau'}{d \tau^{2}} \frac{d x^{\beta}}{d \tau'} + \Big( \frac{d \tau'}{d \tau} \Big)^{2} \frac{ d^{2} x^{\beta}}{d \tau'^{2}}  \right\rbrace 
+  \Gamma^{\eta}_{\alpha \beta} \Gamma^{\beta}_{\rho \sigma}  \Big( \frac{d \tau'}{d \tau} \Big)^{2}   \frac{d x^{\alpha}}{ d \tau'}   \frac{d x^{\rho}}{ d \tau'}   \frac{d x^{\sigma}}{ d \tau'}   
 \Big]  
 \\
+   \frac{q^{2}}{3m} (R^{\mu}_{\lambda} u^{\lambda} + R^{\nu}_{\lambda} u^{\lambda} u^{\mu} u_{\nu} )
 + 
   \frac{2q^{2}}{m} f^{\mu \nu}_{Tail} \frac{d \tau'}{d \tau} \frac{ d x_{\nu}}{d \tau'} + a^{\mu}   .
\end{aligned} 
\end{equation}  
We substitute the expression of the $\frac{ d^{2} x^{\mu}}{d \tau'^{2}} $ in the LHS of the above Eqn.(\ref{1.5}) from the Eqn.(\ref{1.1}) and hence obtain :
\begin{equation}\label{1.6}
\begin{aligned}
\frac{d^{2} \tau'}{d \tau^{2}} \frac{d x^{\mu}}{d \tau'} + \Big( \frac{d \tau'}{d \tau} \Big)^{2} \Big[- \Gamma'^{\mu}_{\nu \rho}\frac{d x^{\nu}}{d \tau'} \frac{d x^{\rho}}{d \tau'} + \frac{q}{m} F'^{\mu}_{\, \nu} u'^{\nu} + 
\frac{2q^{2}}{3m} \Big( \frac{D^{2} u'^{\mu}}{d\tau'^{2}} + u'^{\mu} u'_{\nu} \frac{D^{2} u'^{\nu}}{d \tau'^{2}} \Big)   +
\\   
\frac{q^{2}}{3m} (R'^{\mu}_{\lambda} u'^{\lambda} + R'^{\nu}_{\lambda} u'^{\lambda} u'^{\mu} u'_{\nu} ) +    \frac{2q^{2}}{m} f'^{\mu \nu}_{Tail} u'_{\nu}  \Big]
+   \Gamma^{\mu}_{\nu \rho} \Big(\frac{d \tau'}{d \tau} \Big)^{2} \frac{d x^{\nu}}{d \tau} \frac{d x^{\rho}}{d \tau} =   
\frac{q}{m} F^{\mu}_{\,\, \nu} \frac{d \tau'}{d \tau} \frac{d x^{\nu}}{ d \tau'} +
\\
 \frac{2 q^{2}}{3 m } \Big( g^{\mu}_{\eta} +  \Big(\frac{d \tau'}{d \tau} \Big)^{2}  \frac{d x^{\mu}}{ d \tau'}  \frac{d x_{\eta}}{ d \tau'} \Big)  
\Big[   \left\lbrace \Big( \frac{d^{3}\tau'}{d \tau^{3}} \frac{d \tau'}{d \tau} \Big)  \frac{d x^{\eta}}{d\tau'} + 3\Big( \frac{d \tau'}{d \tau} \frac{d^{2} \tau'}{d \tau^{2}} \Big)  \frac{d^{2} x^{\eta}}{d \tau'^{2}} +  \Big(  \frac{d \tau'}{d \tau}  \Big)^{3} \frac{d^{3} x^{\eta}}{d \tau'^{3}}  \right\rbrace +  
\\
 \Big(  \frac{d \tau'}{d \tau}  \Big)^{3} \partial_{\gamma} \Gamma^{\eta}_{\alpha \beta}   \frac{d x^{\alpha}}{ d \tau'} \frac{d x^{\beta}}{ d \tau'} \frac{d x^{\gamma}}{ d \tau'}  +
  3 \Gamma^{\eta}_{\alpha \beta}  \frac{d \tau'}{d \tau}   \frac{d x^{\alpha}}{d \tau'} \left\lbrace \frac{d^{2} \tau'}{d \tau^{2}} \frac{d x^{\beta}}{d \tau'} + \Big( \frac{d \tau'}{d \tau} \Big)^{2} \frac{ d^{2} x^{\beta}}{d \tau'^{2}}  \right\rbrace 
  \\
+  \Gamma^{\eta}_{\alpha \beta} \Gamma^{\beta}_{\rho \sigma}  \Big( \frac{d \tau'}{d \tau} \Big)^{2}   \frac{d x^{\alpha}}{ d \tau'}   \frac{d x^{\rho}}{ d \tau'}   \frac{d x^{\sigma}}{ d \tau'}   
 \Big]  +  \frac{q^{2}}{3m} (R^{\mu}_{\lambda} u^{\lambda} + R^{\nu}_{\lambda} u^{\lambda} u^{\mu} u_{\nu} )
 +  
   \frac{2q^{2}}{m} f^{\mu \nu}_{Tail} \frac{d \tau'}{d \tau} \frac{ d x_{\nu}}{d \tau'} + a^{\mu}  .
\end{aligned} 
\end{equation}
Now we arrange the equation in such a way that the similar terms in the perturbed and unperturbed metric come together so that it can be identified. Doing so Eqn.(\ref{1.6}) can be written in the following form :
\begin{equation} \label{1.7}
\begin{aligned}
\frac{d^{2} \tau'}{d \tau^{2}} \frac{d x^{\mu}}{d \tau'} + \Big( \frac{d \tau'}{d \tau} \Big)^{2} (- \Gamma'^{\mu}_{\nu \rho} +\Gamma^{\mu}_{\nu \rho}) \frac{ d x^{\nu}}{d \tau'}\frac{ d x^{\rho}}{d \tau'}  +  \Big( \frac{d \tau'}{d \tau} \Big)^{2}  \frac{q}{m} \big( F'^{\mu \nu} - \frac{d \tau}{d \tau'}F^{\mu \nu}   \big) \frac{d x_{\nu}}{d \tau'} + 
 \\
  \Big( \frac{d \tau'}{d \tau} \Big)^{2} \frac{q^{2}}{3 m}  ((R'^{\mu}_{\lambda} u'^{\lambda} + R'^{\nu}_{\lambda} u'^{\lambda} u'^{\mu} u'_{\nu} ) -  \Big(\frac{d \tau'}{d \tau} \Big)^{-2}(R^{\mu}_{\lambda} u^{\lambda} + R^{\nu}_{\lambda} u^{\lambda} u^{\mu} u_{\nu} )) +
  \\
 \frac{2 q^{2}}{m} \Big(\frac{d \tau'}{d \tau} \Big)^{2} (f'^{\mu \nu}_{Tail} -  \frac{d \tau}{d \tau'}  f^{\mu \nu}_{Tail} ) u'_{\nu} +
 \\
\frac{2q^{2}}{3m} \Big(\frac{d \tau'}{d \tau} \Big)^{2} \frac{d^{3} x^{\eta}}{d \tau'^{3}} 
\left\lbrace \Big( g'^{\mu}_{\eta} +   \frac{d x^{\mu}}{ d \tau'}  \frac{d x_{\eta}}{ d \tau'} \Big)   - \frac{d \tau'}{d \tau}  \Big( g^{\mu}_{\eta} +  \Big(\frac{d \tau'}{d \tau} \Big)^{2}  \frac{d x^{\mu}}{ d \tau'}  \frac{d x_{\eta}}{ d \tau'} \Big)  \right\rbrace    + 
\\
\frac{2q^{2}}{3m} \Big(\frac{d \tau'}{d \tau} \Big)^{2}  \frac{ d x^{\alpha}}{d \tau'}\frac{ d x^{\beta}}{d \tau'}\frac{ d x^{\gamma}}{d \tau'} 
\left\lbrace \Big( g'^{\mu}_{\eta} +   \frac{d x^{\mu}}{ d \tau'}  \frac{d x_{\eta}}{ d \tau'} \Big)  \partial_{\gamma} \Gamma'^{\eta}_{\alpha \beta} -  \frac{d \tau'}{d \tau}  \Big( g^{\mu}_{\eta} +  \Big(\frac{d \tau'}{d \tau} \Big)^{2}  \frac{d x^{\mu}}{ d \tau'}  \frac{d x_{\eta}}{ d \tau'} \Big)      \partial_{\gamma} \Gamma^{\eta}_{\alpha \beta}  \right\rbrace   +
\\
\frac{2q^{2}}{3m} \Big(\frac{d \tau'}{d \tau} \Big)^{2}  \frac{ d x^{\alpha}}{d \tau'}\frac{ d^{2} x^{\beta}}{d \tau'^{2}}
\left\lbrace \Big( g'^{\mu}_{\eta} +   \frac{d x^{\mu}}{ d \tau'}  \frac{d x_{\eta}}{ d \tau'} \Big)  3 \Gamma'^{\eta}_{\alpha \beta} -  \frac{d \tau'}{d \tau}  \Big( g^{\mu}_{\eta} +  \Big(\frac{d \tau'}{d \tau} \Big)^{2}  \frac{d x^{\mu}}{ d \tau'}  \frac{d x_{\eta}}{ d \tau'} \Big)   3 \Gamma^{\eta}_{\alpha \beta}  \right\rbrace   +
\\
\frac{2q^{2}}{3m} \Big(\frac{d \tau'}{d \tau} \Big)^{2} \frac{ d x^{\alpha}}{d \tau'}\frac{ d x^{\rho}}{d \tau'}\frac{ d x^{\sigma}}{d \tau'}
  \left\lbrace    \Big( g'^{\mu}_{\eta} +   \frac{d x^{\mu}}{ d \tau'}  \frac{d x_{\eta}}{ d \tau'} \Big)   \Gamma'^{\eta}_{\alpha \beta} \Gamma'^{\beta}_{\rho \sigma} -  \frac{d \tau'}{d \tau}  \Big( g^{\mu}_{\eta} +  \Big(\frac{d \tau'}{d \tau} \Big)^{2}  \frac{d x^{\mu}}{ d \tau'}  \frac{d x_{\eta}}{ d \tau'} \Big)    \Gamma^{\eta}_{\alpha \beta} \Gamma^{\beta}_{\rho \sigma}  \right\rbrace  -
  \\
  \frac{2q^{2}}{3m}  \Big( g'^{\mu}_{\eta} +   \frac{d x^{\mu}}{ d \tau'}  \frac{d x_{\eta}}{ d \tau'} \Big)         \left\lbrace  3 \Gamma^{\eta}_{\alpha \beta}  \frac{d \tau'}{d \tau} \frac{d^{2} \tau'}{d \tau^{2}} \Big(  \frac{d x^{\alpha}}{ d \tau'}\frac{d x^{\beta}}{ d \tau'} \Big) +   \frac{d^{3} \tau'}{d \tau^{3}} \frac{d \tau'}{d \tau} \frac{d x^{\eta}}{d \tau'}  +  3   \frac{d \tau'}{d \tau} \frac{d^{2} \tau'}{d \tau^{2}} \frac{d^{2} x^{\eta}}{d \tau'^{2}}  \right\rbrace 
  = a^{\mu}   .
\end{aligned}
\end{equation}
From now, we shall neglect the terms containing Ricci tensors, due to interaction with surrounding matter. The reason, for which we neglect these terms, is explained in detail in Appendix-2. \\ 
We simplify different terms as differences between quantities in the unperturbed metric i.e. with respect to proper time $ \tau $ and in the perturbed metric i.e. with respect to proper time $ \tau '$ as follows :
\begin{equation}  \label{1.8}
\begin{aligned}
\frac{d^{2} \tau'}{d \tau^{2}} \frac{d x^{\mu}}{d \tau'} + \Big( \frac{d \tau'}{d \tau} \Big)^{2} (- \Delta \Gamma^{\mu}_{\nu \rho}) \frac{ d x^{\nu}}{d \tau'}\frac{ d x^{\rho}}{d \tau'}- \frac{1}{2} q F^{\mu}_{\,\, \nu} u^{\nu} u^{\alpha} u^{\beta} h^{R}_{\alpha \beta} - q (g^{\mu \nu} + u^{\mu} u^{\nu} ) h^{R}_{\nu \alpha} F^{\alpha}_{\,\, \beta } u^{\beta}    
\\  
  + \frac{2q^{2}}{3m} \Big(\frac{d \tau'}{d \tau} \Big)^{2} \frac{d^{3} x^{\eta}}{d \tau'^{3}}  \left\lbrace   (1- \xi_{1}) g^{\mu}_{\eta} + h^{R \,\mu}_{\, \, \eta} + (1 - \xi_{1}^{3}) \frac{ d x^{\mu}}{d \tau'}\frac{ d x_{\eta}}{d \tau'} \right\rbrace +
    \frac{2q^{2}}{3m} \Big(\frac{d \tau'}{d \tau} \Big)^{2} \frac{d x^{\alpha}}{d \tau'}\frac{d x^{\beta}}{d \tau'}\frac{d x^{\gamma}}{d \tau'} 
\\
\left\lbrace   ( 1 -  \xi_{1} ) g^{\mu}_{\eta} \partial_{\gamma} \Gamma^{\eta}_{\alpha \beta} + g^{\mu}_{\eta} \partial_{\gamma} \Delta \Gamma^{\eta}_{\alpha \beta} + h^{R \, \mu}_{\, \, \eta} \partial_{\gamma} \Gamma^{\eta}_{\alpha \beta} + \Big(  (1-\xi_{1}^{3}) \partial_{\gamma}\Gamma^{\eta}_{\alpha \beta} + \partial_{\gamma} \Delta \Gamma^{\eta}_{\alpha \beta} \Big) \frac{d x^{\mu}}{d \tau'}  \frac{d x_{\eta}}{d \tau'}  \right\rbrace  
\\
+ \frac{2q^{2}}{3m} \Big(\frac{d \tau'}{d \tau} \Big)^{2}  \frac{ d x^{\alpha}}{d \tau'}\frac{ d^{2} x^{\beta}}{d \tau'^{2}} 3 \left\lbrace  (1- \xi_{1}) g^{\mu}_{\eta} \Gamma^{\eta}_{\alpha \beta} + h^{R \, \mu}_{\, \, \eta} \Gamma^{\eta}_{\alpha \beta}  + g^{\mu}_{\eta} \Delta \Gamma^{\eta}_{\alpha \beta}  + \frac{d x^{\mu}}{d \tau'}  \frac{d x_{\eta}}{d \tau'} \Big( (1- \xi_{1}^{3}) \Gamma^{\eta}_{\alpha \beta}  + \Delta \Gamma^{\eta}_{\alpha \beta}  \Big)
\right\rbrace 
\\
+ \frac{2q^{2}}{3m} \Big(\frac{d \tau'}{d \tau} \Big)^{2}  \frac{ d x^{\alpha}}{d \tau'}\frac{ d x^{\rho}}{d \tau'}\frac{ d x^{\sigma}}{d \tau'} 
\Big\lbrace  (1- \xi_{1}) g^{\mu}_{\eta} \Gamma^{\eta}_{\alpha \beta} \Gamma^{\beta}_{\rho \sigma } + h^{R \,\mu}_{\, \, \eta} \Gamma^{\eta}_{\alpha \beta} \Gamma^{\beta}_{\rho \sigma } + g^{\mu}_{\eta} (\Delta \Gamma^{\eta}_{\alpha \beta})  \Gamma^{\beta}_{\rho \sigma }  + g^{\mu}_{\eta} \Gamma^{\eta}_{\alpha \beta} (\Delta\Gamma^{\beta}_{\rho \sigma })   
\\
+ \frac{d x^{\mu}}{d \tau'} \frac{d x_{\eta}}{d \tau'} \Big( (1- \xi_{1}^{3}) \Gamma^{\eta}_{\alpha \beta} \Gamma^{\beta}_{\rho \sigma} +  \Gamma^{\beta}_{\rho \sigma} \Delta  \Gamma^{\eta}_{\alpha \beta} +   \Gamma^{\eta}_{\alpha \beta} \Delta \Gamma^{\beta}_{\rho \sigma}  \Big)  \Big\rbrace  +  
 \frac{2 q^{2}}{m} \Big(\frac{d \tau'}{d \tau} \Big)^{2} (f'^{\mu \nu}_{ Tail} -  \frac{d \tau}{d \tau'}  
 f^{\mu \nu}_{ Tail} ) \frac{d x_{\nu}}{d \tau'}  
 \\
 - \frac{2q^{2}}{3m}  \Big( g'^{\mu}_{\eta} +   \frac{d x^{\mu}}{ d \tau'}  \frac{d x_{\eta}}{ d \tau'} \Big)         \left\lbrace  3 \Gamma^{\eta}_{\alpha \beta}  \frac{d \tau'}{d \tau} \frac{d^{2} \tau'}{d \tau^{2}} \Big(  \frac{d x^{\alpha}}{ d \tau'}\frac{d x^{\beta}}{ d \tau'} \Big) +   \frac{d^{3} \tau'}{d \tau^{3}} \frac{d \tau'}{d \tau} \frac{d x^{\eta}}{d \tau'}  +  3   \frac{d \tau'}{d \tau} \frac{d^{2} \tau'}{d \tau^{2}} \frac{d^{2} x^{\eta}}{d \tau'^{2}}  \right\rbrace  = a^{\mu}   \, ,  
\end{aligned}
\end{equation}   
\end{widetext}
where the terms $ - \frac{1}{2} q F^{\mu}_{\, \, \nu} u^{\nu} u^{\alpha} u^{\beta} h^{R}_{\alpha \beta}$ and $ - q (g^{\mu \nu} + u^{\mu} u^{\nu} ) h^{R}_{\nu \alpha} F^{\alpha}_{\,\, \beta } u^{\beta} $ have already been derived by P. Zimmerman and E. Poisson in an earlier work \cite{Zimmerman_et_al}.\\ 
For brevity, we introduced the quantity
\begin{equation}
 \xi_{1} = \frac{d \tau'}{d \tau}   \, .
 \end{equation}
In the above equation the quantity $\Delta \Gamma^{\mu}_{\nu \rho} $ is the part of the perturbation in the Christoffel symbol tensor, caused by the regular part $h^{R}_{\mu \nu} $ of the metric perturbation and is given by :
\begin{equation} \label{1.9}
\begin{aligned}
\Delta \Gamma^{\mu}_{\nu \rho} = \Gamma'^{\mu}_{\nu \rho} - \Gamma^{\mu}_{\nu \rho} = \frac{1}{2} h^{R \, \mu \alpha} (\partial_{\rho} g_{\alpha \nu} + \partial_{\nu} g_{\alpha \rho} - \partial_{\alpha} g_{\nu \lambda} ) + 
\\
\frac{1}{2} g^{\mu \alpha} (\partial_{\rho} h^{R}_{\alpha \nu} + \partial_{\nu} h^{R}_{\alpha \rho} - \partial_{\alpha} h^{R}_{\nu \lambda} )   .
\end{aligned}
\end{equation} 
The above expression of $ \Delta \Gamma^{\mu}_{\nu \rho}$ can be further simplified as :
\begin{equation} \label{1.10}
\Delta \Gamma^{\mu}_{\nu \rho} = \frac{1}{2} g^{\mu \alpha} (\nabla_{\rho} h^{R}_{\alpha \nu} + \nabla_{\nu} h^{R}_{\alpha \rho} - \nabla_{\alpha} h^{R}_{\nu \lambda} )    .
\end{equation}
%____________________________________________________________________________________________________________
%____________________________________________________________________________________________________________
\section{Extra terms generated due to perturbation of the electromagnetic self-force and their significance :}  
In this section, we investigate the significance of the additional perturbative terms generated due to metric perturbations from the electromagnetic self-force in comparison with the gravitational self-force. The gravitational radiation reaction on the motion of the particle in curved space time in the absence of electromagnetic radiation reaction is given by :
\begin{equation} \label{4.1}
\begin{aligned}
{\displaystyle a^{\mu}_{1 } =  \frac{d^{2} \tau'}{d \tau^{2}} \frac{d \tau}{d \tau'} u^{\mu}-
  \Delta \Gamma^{\mu}_{\nu \rho} u^{\nu} u^{\rho}  } 
  = -\xi_{1}^{2} (\delta^{\mu}_{\eta} + u^{\mu}u_{\eta}) \Delta \Gamma^{\eta}_{\alpha \beta } u'^{\alpha} u'^{\beta} .
\end{aligned}
\end{equation}
It is to be noted that the above simplified expression of the gravitational radiation reaction term can be obtained by applying the orthogonality property of the reaction, in the case where there is no electromagnetic radiation reaction \ \cite{Barack}.\\ 
Now, the additional perturbative terms, in linear order of the metric perturbation $h^{R}_{\alpha \beta}$, generated from the electromagnetic radiation reaction due to the metric perturbations emitted from the particle, which are absent when there is only one among electromagnetic self-force and metric-perturbations, are : 
\begin{equation} \label{4.1a}
a^{\mu }_{int \, 1} = - \frac{1}{2} \frac{q}{m} F^{\mu}_{\,\, \nu} u^{\nu} u^{\alpha} u^{\beta} h^{R}_{\alpha \beta} \, , 
\end{equation}
\begin{equation} \label{4.1b}
a^{\mu }_{int \, 2} = - \frac{q}{m} (g^{\mu \nu} + u^{\mu} u^{\nu} ) h^{R}_{\nu \alpha} F^{\alpha}_{\,\, \beta } u^{\beta} \, , 
\end{equation}
\begin{equation} \label{4.2}
a^{\mu}_{2} = \frac{2 q^{2}}{3 m } \xi_{1}^{2} h^{R \, \mu}_{\,\, \eta} \frac{d^{2} u'^{\eta}}{d \tau'^{2}}  \, , 
\end{equation}
\begin{equation} \label{4.3}
a^{\mu}_{3 } =   \frac{2 q^{2}}{ m } \xi_{1}^{2} u'^{\alpha} \frac{d u'^{\beta}}{ d \tau'} (h^{R \, \mu}_{\, \, \eta} \Gamma^{\eta}_{\alpha \beta} + \Delta \Gamma^{\mu}_{\alpha \beta} + \Delta \Gamma^{\eta}_{\alpha \beta} u'^{\mu} u'_{\eta})  \, , 
\end{equation}
\begin{equation} \label{4.4}
\begin{aligned}
a^{\mu}_{4 } =   \frac{2 q^{2}}{ 3 m } \xi_{1}^{2} u'^{\alpha} u'^{\beta} u'^{\gamma} ( \partial_{\gamma} \Delta\Gamma^{\mu}_{\alpha \beta} + h^{R \, \mu}_{\,\, \eta} \partial_{\gamma} \Gamma^{\eta}_{\alpha \beta} +
 u'^{\mu} u'_{\eta} \partial_{\gamma} \Delta\Gamma^{\eta}_{\alpha \beta} )  \,  ,
\end{aligned}
\end{equation}
\begin{center}
and 
\end{center}
\begin{equation} \label{4.5}
\begin{aligned}
a^{\mu}_{5 } =   \frac{2 q^{2}}{ 3 m } \xi_{1}^{2} u'^{\alpha} u'^{\rho} u'^{\sigma} (h^{R \, \mu}_{\,\, \eta} \Gamma^{\eta}_{\alpha \beta} \Gamma^{\beta}_{ \rho \sigma} + \Gamma^{\beta}_{ \rho \sigma} \Delta \Gamma^{\mu}_{\alpha \beta} + \\
\Gamma^{\mu}_{\alpha \beta} \Delta  \Gamma^{\beta}_{ \rho \sigma} + u'^{\mu} u'_{\eta} (\Gamma^{\beta}_{ \rho \sigma} \Delta \Gamma^{\eta}_{\alpha \beta} + \Gamma^{\eta}_{\alpha \beta} \Delta  \Gamma^{\beta}_{ \rho \sigma}  ))  .
\end{aligned}
\end{equation}
%Some corrections and additions are done 
In this case, there may be a confusion that why we have written these correction terms in equations \ref{4.2} to \ref{4.5} separately, although their basic-source is same : the second term on the R.H.S. of the equation \ref{1.1} i.e. the Abraham-Lorentz-Dirac term. The simple reason behind this is that despite having identical source, these terms originate from four different kind of terms within the Abraham-Lorentz-Dirac term. This can be clearly checked from the equation \ref{1.2.2}, where we have written the detailed expression of $\frac{D^{2} u^{\mu}}{d \tau^{2}} $, expanding the covarinat-derivatives within it. \\
In the next subsections we analyze the significance of the terms $a^{\mu}_{3}, \, a^{\mu}_{4} $ and $ a^{\mu}_{5 } $ with respect to the gravitational radiation reaction term $ a^{\mu}_{1 } $. We shall avoid discussing the significance of the term $ a^{\mu}_{2 }$, as it contains time-rate of change of acceleration and hence this is very complicated to compare for practical astrophysical and cosmological phenomena. 
%------------------------------------------------------------------------------
%------------------------------------------------------------------------------
\subsection{The significance of the term $ a^{\mu }_{3} $ : }
Let us now  analyze the ratio :
\begin{equation} \label{4.6}
  \frac{a^{\mu}_{3 }}{a^{\mu}_{1}} = \frac{ {\displaystyle \frac{2 q^{2}}{ m } \xi_{1}^{2} u'^{\alpha} \frac{d u'^{\beta}}{ d \tau'} (h^{R \, \mu}_{\, \, \eta} \Gamma^{\eta}_{\alpha \beta} + \Delta \Gamma^{\mu}_{\alpha \beta} + \Delta \Gamma^{\eta}_{\alpha \beta} u'^{\mu} u'_{\eta}) }}
  { {\displaystyle -\xi_{1}^{2} (\delta^{\mu}_{\eta} + u^{\mu}u_{\eta}) \Delta \Gamma^{\eta}_{\alpha \beta } u'^{\alpha} u'^{\beta} }}   .  
\end{equation}
We see that for $ a^{\mu}_{3 }  \sim a^{\mu}_{1 }$, one of the requirement is:
%\begin{equation} \label{4.7a}
%\frac{2 q^{2}}{ m } \frac{d u'^{\beta}}{ d \tau'}  \sim  u'^{\beta}
%\end{equation} 
%and 
\begin{equation} \label{4.7b}
\frac{2 q^{2}}{ m } \xi_{1}^{2} u'^{\alpha} \frac{d u'^{\beta}}{ d \tau'} \Delta \Gamma^{\eta}_{\alpha \beta} u'^{\mu} u'_{\eta}  \sim     \xi_{1}^{2} u'^{\alpha} \Delta \Gamma^{\eta}_{\alpha \beta}  u'^{\beta}  u^{\mu} u_{\eta}   . 
\end{equation}
After substituting $ u^{\mu} u_{\eta} \equiv \Big( \frac{d\tau'}{d\tau} \Big)^{2}  u'^{\mu} u'_{\eta} $ and cancelling out the $u'^{\mu } $ from both sides of the above condition, we obtain : 
\begin{equation} \label{4.7c}
{\displaystyle  \frac{2 q^{2}}{ m } \xi_{1}^{2} u'^{\alpha} \frac{d u'^{\beta}}{ d \tau'} \Delta \Gamma^{\eta}_{\alpha \beta} u'_{\eta} \sim  \xi_{1}^{4} u'^{\alpha} \Delta \Gamma^{\eta}_{\alpha \beta}  u'^{\beta} u'_{\eta}  }  .
\end{equation} 
It is to be noticed that the above relation is a tensorial one where on  both sides three indices $\alpha, \beta $ and $ \eta $  are repeated indices and they are contracted among the tensors in such a way that we can not cancel out the term $ \Delta \Gamma^{\eta}_{\alpha \beta} u'^{\alpha} u'_{\eta} $ from both sides of Eqn.(\ref{4.7c}), although that is common.  For that,  we write the expanded expression of the quantity ${\displaystyle  \Delta \Gamma^{\eta}_{\alpha \beta}          \frac{d u'^{\beta} }{d \tau'} u'^{\alpha} u'_{\eta}   }$ with respect to the repeated index $\beta $ :  
\begin{center}
${\displaystyle \Delta \Gamma^{\eta}_{\alpha \beta} \frac{d u'^{\beta} }{d \tau'} u'^{\alpha} u'_{\eta} =   
 \Delta \Gamma^{\eta}_{\alpha r} \frac{d u'^{r} }{d \tau'} u'^{\alpha} u'_{\eta} + \Delta \Gamma^{\eta}_{\alpha \theta} \frac{d u'^{\theta} }{d \tau'} u'^{\alpha} u'_{\eta} +
 }$ \\
 ${\displaystyle
  \Delta \Gamma^{\eta}_{\alpha \phi} \frac{d u'^{\phi} }{d \tau'} u'^{\alpha} u'_{\eta} +  \Delta \Gamma^{\eta}_{\alpha t } \frac{d u'^{t } }{d \tau'} u'^{\alpha} u'_{\eta}   }$ .
\end{center}
Where the indices $r, \theta, \phi $ and $t $ denote the four coordinates of the coordinate system. In a similar way, we expand the quantity $ \Delta \Gamma^{\eta}_{\alpha \beta}  u'^{\beta} u'_{\eta} u'^{\alpha}$ with respect to $\beta$, and then substituting the expanded forms of these two quantities in both sides of the relation (\ref{4.7c}) to obtain the % with respect to $\beta $, in both sides of the relation \ref{4.7c}, then we may say that 
condition to be satisfied in coordinate-wise manner, we get
\footnote{Although the set of conditions (\ref{4.7d}) to (\ref{4.7g}) together obviously satisfy the condition (\ref{4.7c}), it is not necessarily the only case for satisfying this condition, as the terms in the above equations are summed up in (\ref{4.7c}). We take the condition in coordinate-wise for brevity of our analysis.} : 
\begin{eqnarray} \label{4.7d}
\frac{2 q^{2}}{ m } \xi_{1}^{2} u'^{\alpha} \frac{d u'^{r}}{ d \tau'} \Delta \Gamma^{\eta}_{\alpha r} u'_{\eta} \sim  \xi_{1}^{4} u'^{\alpha} \Delta \Gamma^{\eta}_{\alpha r}  u'^{r} u'_{\eta} ,
\\
\label{4.7e}
\frac{2 q^{2}}{ m } \xi_{1}^{2} u'^{\alpha} \frac{d u'^{\theta}}{ d \tau'} \Delta \Gamma^{\eta}_{\alpha \theta} u'_{\eta} \sim  \xi_{1}^{4} u'^{\alpha} \Delta \Gamma^{\eta}_{\alpha \theta}  u'^{\theta} u'_{\eta} ,
\\
\label{4.7f}
\frac{2 q^{2}}{ m } \xi_{1}^{2} u'^{\alpha} \frac{d u'^{\phi}}{ d \tau'} \Delta \Gamma^{\eta}_{\alpha \phi} u'_{\eta} \sim  \xi_{1}^{4} u'^{\alpha} \Delta \Gamma^{\eta}_{\alpha \phi}  u'^{\phi} u'_{\eta}  , 
\\
\label{4.7g}
\frac{2 q^{2}}{ m } \xi_{1}^{2} u'^{\alpha} \frac{d u'^{t}}{ d \tau'} \Delta \Gamma^{\eta}_{\alpha t} u'_{\eta} \sim  \xi_{1}^{4} u'^{\alpha} \Delta \Gamma^{\eta}_{\alpha t}  u'^{t} u'_{\eta}   .
\end{eqnarray} 
Let us consider any one of  the above set of relations, say the first one based on radial coordinate r. It gives : 
\begin{equation} \label{4.8}
\frac{2 q^{2}}{ m }  \frac{d u'^{r}}{ d \tau'}  \sim  \xi_{1}^{2}   u'^{r}  ;
\end{equation}
provided that the quantity $ \Delta \Gamma^{\eta}_{\alpha r} u'_{\eta}  u'^{\alpha} $ is non-zero, which should be obvious for any curved space time as any of the spatial components of the four-velocity must be non-zero for the motion of the particle or compact object, as well as any of the perturbed component of the Christoffel.  
%\vspace{1cm}\\
%Actually both the conditions are same as in this case $ \xi_{1} \sim 1 $. 
%\\
Thus, the condition (\ref{4.8}) in S.I. units reads:\footnote{As, we have considered the metric perturbation to be sufficiently small, so that the perturbation terms are retained up to linear order only ; hence the difference between the corresponding proper times of perturbed and unperturbed metrics viz. $\tau' $ and $\tau $ should also be small enough such that $\frac{d \tau'}{d \tau } \, \sim \, 1 $ ).}
\begin{equation} \label{4.9}
\frac{q^{2}}{2 \pi \epsilon_{0} c^{3} m }  \frac{d u'^{r} }{ d \tau'}  \sim  u'^{r}  .
\end{equation}
We describe the fulfillment of the condition (\ref{4.9}) in two different classes of charged objects : (i)  charged sub-atomic particles (e.g. we  estimate it numerically for a  proton) and (ii) charged neutron stars or stellar mass black holes. 
If we consider  the orbiting particle to  be a proton, the above condition yields % having magnitude of charge $ |q| \, \approx 1.6 \times 10^{-19} \, C $ and mass $m \, \approx 1.67 \times 10^{-27} \, kg$. Then the above condition gives :
%\begin{equation} \label{4.10}
%\begin{aligned}
%\frac{2 \times 8.99 \times 10^{9}}{(3 \times 10^{8})^{3}} \times \frac{(1.6 \times 10^{-19})^{2}}{1.67 \times 10^{-27} } \, \frac{(Nm^{2}kg^{-1})}{m^{3}s^{-3}} \frac{d u'^{\beta}}{ d \tau'} \, (in \, ms^{-2}) \, \\  \sim
 %\,  u'^{\beta}  (in \, ms^{-1} )
%\end{aligned}
%\end{equation}
%which gives :
%\begin{equation} 
%10^{-26} (in \, s)  \frac{d u'^{r}}{ d \tau'} (in \, ms^{-2})  \sim  u'^{r} (in \, ms^{-1}) 
%\end{equation}
%or, 
\begin{equation} \label{4.10}
\frac{ {\displaystyle \Big( \frac{d u'^{r}}{ d \tau'} \Big) } }{  u'^{r}  } \sim \,  10^{26} \, s^{-1}    .
\end{equation}
Therefore, for a proton orbiting around a black hole the condition  for the significance  of the perturbative term $ a^{\mu}_{3 } $ is that the acceleration of the proton has to be $ 10^{26} $ order larger than the speed\footnote{Here we use proton  just  to get an idea on the order of acceleration required for such sub-atomic particles to satisfy the condition \ref{4.9}.  }.\\
It would be  interesting to test the significance of the term $a^{\mu}_{3} $ in the cases of charged stars, specially charged neutron stars or white dwarfs, or charged stellar mass black holes revolving around a supermassive black hole. Although, still there is no distinct astronomical evidence for any compact object containing a significant amount of net electric charge, specially for stars and black holes, many researchers have been working on theoretical models of stars containing a significant amount of net electric charge. For instance, in the reference~\cite{Kumar_et_al} the authors have discussed a class of static stellar equilibrium configurations of relativistic  spheres made of charged perfect fluids, where they have analyzed the physical acceptability of their theoretical model for some compact star candidates like SAX J1808.4-3658, 4U 1538-52, PSR J1903+327, Vela X-1 and 4U1608-52. They have concluded that their results strongly suggest that a class of compact stellar models with charged perfect fluid matter distribution is permitted with the new solution discussed in their work. \\

According to the references ~\cite{Ray_et_al, Ghezzi_et_al, Varela_et_al}, and \cite{Ray2_et_al}, the global balance of forces allows a net charge as large as $10^{20} \, C$ in neutron stars, producing a very high electric field of order $\sim \, 10^{21} \, V/m$.
% \textcolor{red}{Then,  for a neutron star of mass }
Then the condition (\ref{4.9}) for that would be : 
\begin{eqnarray}  \label{4.11}
%${\displaystyle \frac{2 \times 8.99 \times 10^{9}}{(3 \times 10^{8})^{3}} \times \frac{(10^{20})^{2}}{ k \times 2 \times 10^{30} } \frac{d u'^{\beta}}{d \tau'}  \sim u'^{\beta}   }$ \\
%${\displaystyle  
 \frac{d u'^{\beta}}{d \tau'}  \sim  \Big( \frac{M_{NS}}{M_{\odot}} \times 10^{6} \, s^{-1}\Big) \,  u'^{\beta}    ,
 % }$
\end{eqnarray}
where $M_{NS}$ and $M_{\odot}$ denote the mass of the neutron star and the solar mass, respectively. We know that usually neutron stars and white dwarfs have mass of the order of solar mass $M_{\odot} $ i.e. $ M_{NS}/ M_{\odot} \sim 1$. 
Here, we consider the observational work in the reference~\cite{GRAVITY_Collaboration}, which reports that the speed of a star around the supermassive black hole at the center of Milky-way reaches approximately $ 8 \times 10^{6} \, ms^{-1} $, and hence, in such cases, for satisfying the condition derived above, the acceleration of the star around the supermassive black hole has to be $\sim \, 10^{12} \, ms^{-2} $.
 So, achieving acceleration of this order would be quite difficult if a single neutron star or stellar mass black hole revolves around a supermassive black hole. To satisfy the condition (\ref{4.11}), a different type of astrophysical configuration is required. When a stellar mass black hole binary or a neutron star binary or a neutron star-black hole binary would be revolving around a supermassive black hole, then the smaller components in binary formation within the three-body system should achieve the required acceleration to satisfy the condition (\ref{4.11}). It has been shown in the work by Xian Chen et al \cite{Chen_et_al}, that such EMRIs are expected to be produced by tidal capture of smaller binaries by a supermassive black hole. Again, at the late inspiral stage or at the merging stage, this binary becomes sufficiently compact with respect to the supermassive black hole such that it can be treated with the point-particle equations in our work.  \\
 
 In support of the fact that the smaller components of these EMRIs can achieve such order of acceleration required for satisfaction of the condition (\ref{4.11}), we give an example of the acceleration in the binary black hole candidate GW150914, from which first direct detection of gravitational waves by aLIGO had been done \cite{aLIGO}. For this candidate, during $0.2$ second time interval of the detectable gravitational wave signal, the relative orbiting velocity of the black holes increased from $30\%$ to $60\%$  of the speed of light, and  hence in this case the order of acceleration was approximately $10^{8} \, m. s^{-2} $. Although the merger-stage dynamics of the black holes can be accurately determined by numerical general relativistic techniques only, yet from this estimation of acceleration, we get an intuitive idea that in case of typical binaries of stellar mass black holes or neutron stars or binaries of black hole-neutron star, the acceleration achieved in the late inspiral stage and merger stage would be of similar order or even more.
So, if the smaller component of an EMRI be such a binary, where the stars or stellar mass black holes contain sufficient net electric charge, then satisfying the condition (\ref{4.11}) is clearly possible.
 \\
 
%possible as the star would come closer to the supermassive black hole.  
% IMPORTANT : In this case it would be very good if we can provide calculation of the surface-gravity of a typical supermassive black hole. 
Therefore, binary formations of neutron stars or stellar mass black holes containing a net electric charge of order $ 10^{20} \, C$ and inspiralling around supermassive black holes, are expected to satisfy the condition of 
significance (\ref{4.11}). This type of extreme mass-ratio inspirals (EMRIs) are expected to be detected by the upcoming space-based gravitational wave detector LISA. Hence, if the smaller mass components of such EMRIs contain a significant amount of net charge, then neglecting the term $ a^{\mu }_{3} $, generated due to perturbation of electromagnetic radiation reaction by metric fluctuations or gravitational radiation, may lead to theoretically wrong estimation of the parameters related to these sources of gravitational waves.\\

Next, we consider a period of  early universe  where primordial black holes (PBHs) are  expected to be produced by direct gravitational collapse of sufficiently deep density perturbations. 
Furthermore, during this epoch, as the universe was full of charged particles (i.e. atom formation did not start yet), it is unlikely that the PBHs would be neutral. Therefore, in this case charged particles would inspiral around charged PBHs and ultimately fall into the PBHs, emitting  gravitational and electromagnetic radiation.
%as during the corresponding early era of the Universe, to say in a more specific way, before formation of atoms by binding of eletrons with nuclei, the Universe was full of charged particles, it is highly unlikely that the PBHs %would be neutal. So, in the specified era, there would be systems where charged particles would inspiral around charged PBHs and ultimately fell into the PBHs with emission of gravitational and electromagnetic radiation. 
The interesting fact here is that many of these systems could have the size of atoms. For instance, the PBHs which had mass smaller than  $\sim 10^{20}\, kg $, their  Schwarzschild radius would be  less than $10^{-7} \, m$, and thus, we expect that they can be treated as quantum particles. Therefore, in  such  systems we expect that the charged particles orbiting around charged PBHs would have huge acceleration, as required by the condition (\ref{4.10}), and consequently, the perturbative term $ a^{\mu}_{3} $ is expected to be significant\footnote{An example of a system where we can have a huge acceleration is the revolving of an electron around a nucleus in the Bohr-model of atom. The order of  magnitude of such acceleration is $ \sim \, 10^{22} \, ms^{-2}$. Hence, as in the early Universe the charged particles revolving around charged PBHs constituted systems of atomic-size, emitting both gravitational and electromagnetic radiation, there also the acceleration of the revolving particle would be quite similar, even expected to be larger due to the curvature of the PBH in comparison with nucleus of a typical atom.  }. \\

Another astrophysical phenomenon where the condition (\ref{4.10}) may be satisfied is in `Relativistic Astrophysical Jets'. In these case, accelerated ionized matter are emitted in the form of a beam from some high-energy astrophysical sources and usually the magnitude of their acceleration is huge. If in any of such astrophysical jet, the ions are accelerated during sufficiently small time to relativistic speeds, then the condition (\ref{4.10}) is expected to be satisfied in the part of the jet closest to the source, specially a supermassive black hole at the center of an active galaxy. Even if the accelerated ions of such a relativistic astrophysical jet passes through the vicinity of another black hole or compact object,  we may  also  expect that the condition (\ref{4.10}) to  be satisfied.\footnote{In any  astrophysical scenario,  if  the  the Plasma acceleration of ions can be achieved, then that would be an ideal case for satisfying the  condition (\ref{4.10}). Indeed, in plasma acceleration of ions, the magnitude of acceleration as high as $10^{22}- 10^{23} \, ms^{-2} $ can be reached \cite{Assmann, Rosenzweig_et_al}.}   \\
\vspace{.3cm}\\
%----------- Addition -----------------------------
For  $ a^{\mu}_{3} \sim  a^{\mu}_{1} $, another requirement is:
 %  can be derived, is given by :
\begin{equation} \label{4a}
\frac{2 q^{2}}{ m } \xi_{1}^{2} u'^{\alpha} \frac{d u'^{\beta}}{ d \tau'} h^{R \, \mu}_{\, \, \eta} \Gamma^{\eta}_{\alpha \beta} 
\sim \xi_{1}^{2}    u'^{\alpha}   u'^{\beta}  \Delta \Gamma^{\mu}_{\alpha \beta}     .
\end{equation}
%Expanding the sum over \textcolor{red}{the} repeated index $ \alpha $, 
Analyzing the condition in coordinate-wise manner as we did before ,
%\textcolor{red}{and  considering the first terms},  i.e. when $ \alpha = %r$, we get :
%\begin{equation} \label{4b}
%\frac{2 q^{2}}{ m }  u'^{r} \frac{d u'^{\beta}}{ d \tau'} h^{\mu}_{\eta} \Gamma^{\eta}_{r \beta} 
%\sim   u'^{r}   u'^{\beta}  \Delta \Gamma^{\mu}_{r \beta}    
%\end{equation} 
%After cancelling $u'^{r} $ from both sides of \ref{4b} , 
we obtain for the radial coordinate:
\begin{equation} \label{4c}
\frac{2 q^{2}}{ m }  \frac{d u'^{\beta}}{ d \tau'} h^{R \, \mu}_{\, \, \eta} \Gamma^{\eta}_{r \beta} 
\sim    u'^{\beta}  \Delta \Gamma^{\mu}_{r \beta}      .
\end{equation}
%After using  the expression (\ref{1.9}),   and keep the leading order in the expansion, we get
At this point,  we make use  of Eqn.(\ref{1.9}), and  substitute  just the first term in the expression of  $ \Delta \Gamma^{\mu}_{r \beta} $  into  Eqn.(\ref{4c}), and we get 
% using   the first part of the expression of  $ \Delta \Gamma^{\mu}_{r \beta} $  in  % of $ \Delta \Gamma^{\mu}_{r \beta} $  in (\ref{1.9})%as given earlier viz. :  
%\begin{equation} \label{4d}
%\begin{aligned}
%\Delta \Gamma^{\mu}_{r \beta} = \frac{1}{2} h^{\mu \kappa} (\partial_{r} g_{\kappa \beta} + \partial_{\beta} g_{\kappa r} - \partial_{\kappa} g_{r \beta } ) + 
%\\
%\frac{1}{2} g^{\mu \kappa} (\partial_{r} h_{\kappa \beta} + \partial_{\beta} h_{\kappa r} - \partial_{\kappa} h_{r \beta} )
%\end{aligned}
%\end{equation} 
%At first we use the first part of the $ \Delta \Gamma^{\mu}_{r \beta} $ in the condition \ref{4c} and we also substitute the expression of $ \Gamma^{\eta}_{r \beta}  = \frac{1}{2} g^{\eta \kappa} (\partial_{r} g_{\kappa \beta} + %\partial_{\beta} g_{\kappa r} - \partial_{\kappa} g_{r \beta }) $ getting :
%\begin{equation} \label{4e}
%\begin{aligned}
%\frac{2 q^{2}}{ m } \frac{1}{2}  \frac{d u'^{\beta}}{ d \tau'} h^{\mu}_{\eta}  g^{\eta \kappa} (\partial_{r} g_{\kappa \beta} + \partial_{\beta} g_{\kappa r} - \partial_{\kappa} g_{r \beta })  \sim 
%\\
% u'^{\beta}   \frac{1}{2} h^{\mu \kappa} (\partial_{r} g_{\kappa \beta} + \partial_{\beta} g_{\kappa r} - \partial_{\kappa} g_{r \beta } )  
 %\end{aligned}
%\end{equation} 
% Cancelling out the quantity $h^{\mu \kappa} (\partial_{r} g_{\kappa \beta} + \partial_{\beta} g_{\kappa r} - \partial_{\kappa} g_{r \beta } )  $ from both sides of the conditon \ref{4e}, we obtain :
 \begin{equation} \label{4f}
 \begin{aligned}
\frac{2 q^{2}}{ m }   \frac{d u'^{\beta}}{ d \tau'}  \sim 
 u'^{\beta}     .
 \end{aligned} 
 \end{equation}
It is interesting to note that finally we obtain the identical condition given in Eqn.(\ref{4.8}) (or, equivalently,   in Eqn.(\ref{4.9}) ), and hence the practical cases where this would be satisfied are also same. \\

Next, we compare the left hand side of Eqn.(\ref{4c}) with the part of $ \Delta \Gamma^{\mu}_{r \beta} $  involving derivatives of $h^{R}_{\mu \nu}$, and  we find %obtaining :
\begin{equation} \label{4g}
\begin{aligned}
\frac{2 q^{2}}{ m }   \frac{d u'^{\beta}}{ d \tau'} h^{R \, \mu \kappa} (\partial_{r} g_{\kappa \beta} + \partial_{\beta} g_{\kappa r} - \partial_{\kappa} g_{r \beta })  \sim 
\\
 u'^{\beta}  g^{\mu \kappa} (\partial_{r} h^{R}_{\kappa \beta} + \partial_{\beta} h^{R}_{\kappa r} - \partial_{\kappa} h^{R}_{r \beta } )  \,    .
 \end{aligned} 
\end{equation} 
%Or, this can be written as 
%\textcolor{red}{I  omitted the  next equation, I don't like we express it as a ratio, since these are tensor quantities} 

%\begin{equation} \label{4h}
%\frac{ u'^{\beta}  g^{\mu \kappa} (\partial_{r} h_{\kappa \beta} + \partial_{\beta} h_{\kappa r} - \partial_{\kappa} h_{r \beta } )  }{ {\displaystyle  \frac{d u'^{\beta}}{ d \tau'} h^{\mu \kappa} (\partial_{r} g_{\kappa \beta} + \partial_{\beta} %g_{\kappa r} - \partial_{\kappa} g_{r \beta })   }} \sim \frac{2 q^{2}}{ m }  
%\end{equation}
It is to be noted that whether the above condition would be satisfied in any case, would depend on the associated components of the regular part of gravitational radiation $h^{R \, \mu \kappa} $ and metric $g^{\mu \kappa} $ ( more specifically saying it would depend on the index $\mu $, as the rest of the indices are repeated indices). Here, we are giving a qualitative discussion on satisfying the condition, rather than a quantitative analysis.
% The gravitational radiation varies as $1/r $ from the source, where r is the radial distance from the source, and hence the radial derivative term $ \partial_{r} h_{\kappa \beta} $ would give just $-\frac{1}{r }h_{\kappa \beta} $.
Actually it depends on the fact that how does the regular part of the gravitational radiation
vary with radial distance from the source. 
On the other hand, if we consider spherically symmetric metrics, then the partial derivative with respect to r of $g_{\theta \theta} $ and $g_{\phi \phi} $ would give a factor of $2r$, while those of $g_{tt} $ and $ g_{rr}$ would depend on the particular type of that metric. However, it is well known that if the radial distance r is not too small, then the amplitude of gravitational radiation (and the regular part of this gravitational radiation too) acting here as the metric perturbation, is very much less than that of the metric components. Hence, for  typical  astrophysical systems, the values of quantities $h^{R \, \mu \kappa} $ and $ (\partial_{r} h^{R}_{ \kappa \beta} + \partial_{\beta} h^{R}_{\kappa r} - \partial_{\kappa} h^{R}_{r \beta } )   $ should be very small. While, the values of quantities $g^{\mu \kappa} $ and $ (\partial_{r} g_{\kappa \beta} + \partial_{\beta} g_{\kappa r} - \partial_{\kappa}g_{r \beta } )   $ are relatively very large than them. Hence, depending on the situation it is possible to get some cases where the condition (\ref{4g}) is satisfied.  
\vspace{0.5cm}
\subsection{The significance of the term $ a^{\mu}_{4} $ : }
Next, we take the ratio of the perturbative term $a^{\mu}_{4} $ with the gravitational radiation reaction term $ a^{\mu}_{1}$:
\begin{equation} \label{4.12}
\begin{aligned}
\frac{a^{\mu}_{4}}{a^{\mu}_{1}}  &= 
& \frac{\frac{2 q^{2}}{ 3 m } \xi_{1}^{2} u'^{\alpha} u'^{\beta} u'^{\gamma} ( \partial_{\gamma} \Delta\Gamma^{\mu}_{\alpha \beta} + h^{R \, \mu}_{\eta} \partial_{\gamma} \Gamma^{\eta}_{\alpha \beta} +
 u'^{\mu} u'_{\eta} \partial_{\gamma} \Delta\Gamma^{\eta}_{\alpha \beta} ) }{ {\displaystyle -\xi_{1}^{2} (\delta^{\mu}_{\eta} + u^{\mu}u_{\eta}) \Delta \Gamma^{\eta}_{\alpha \beta } u'^{\alpha} u'^{\beta} }}   ,
\end{aligned}   
\end{equation} 
and one of the conditions for $a^{\mu}_{4} \, \sim  a^{\mu}_{1} $ reads
\begin{equation} \label{4.13}
\frac{2 q^{2}}{ 3 m }  u'^{\alpha} u'^{\beta} u'^{\gamma}  \partial_{\gamma} \Delta\Gamma^{\mu}_{\alpha \beta}  \sim \Delta\Gamma^{\mu}_{\alpha \beta} u'^{\alpha} u'^{\beta}    .
\end{equation}
%----------------------2nd case of correction---------------------------------------
Following the same steps as  in the previous subsection, we find %   follow  similar in case of this condition too the similar argument, which has been stated for the condition \ref{4.7c}, is true . The indices $\alpha, \beta, \gamma $ are repeated indices here, which are contracted among the tensors. In this case also, we can not separate the common part just like the case of the condition \ref{4.7c}. Before proceeding, we insert the expression of perturbation in the Christoffel symbol, given by equation \ref{1.10}.
\begin{equation} \label{4.17}
\frac{2 q^{2}}{ 3 m }    u'^{\gamma} \partial_{\gamma}   g^{\mu r}  \sim  g^{\mu r}    .
\end{equation}
%(We can now cancel out the term $  u'^{\alpha} u'^{\beta} \mathcal{H}_{r \alpha \beta}  $ from both sides because it no longer contains any repeated index, contracted with the rest part .)\\
If we consider the metric to be Reissner-Nordstrom metric, then as the metric is spherically symmetric and time-independent, the above condition (\ref{4.17}) yields :
\begin{equation} \label{4.18}
\frac{2 q^{2}}{ 3 m } ( u'^{r}   \partial_{r} + u'^{\theta} \partial_{\theta}  ) g^{\mu r} \sim g^{\mu r }    .
\end{equation}
It is very interesting to find that not only for $g^{rr}$, but for all non-zero components of the Reissner-Nordstrom metric, the above condition gives in S.I. units:
\begin{equation} \label{4.19}
\frac{q^{2}}{6 \pi \epsilon_{0} c^{3} m } | u'^{r}|  \sim \, r      ,
\end{equation}
where r is the radial distance from the black hole. For protons, the above condition yields 
\begin{equation} \label{4.20}
| u'^{r}| \sim 10^{26} s^{-1} \, r (in \, m) .
\end{equation}
Therefore, we expect the term $ a^{\mu}_{4} $ to be significant for ultra-relativistic motion of protons  around PBHs of  mass less than $10^{9} \, kg$, which is equivalent to Schwarzschild length scale  of $10^{-18} \, m$.
% Note that for the above condition (\ref{4.20}) implies that the radial distance  r  must be smaller than  $10^{-18}~m$ so that  the speed  of the particle does not exceed the speed of light.
% there is  limiting  radial distance r so as any speed can not exceed c i.e. $ 10^{8} \, ms^{-1}$.} \\
%Therefore, for neutron stars or stellar mass black holes containing a net electric charge of order $ 10^{20} \, C$ and \textcolor{red}{inspiralling: please check spelling} around supermassive black holes, are expected to satisfy the condition of significance \textcolor{red}{(\ref{4.8})" PLEASE CHECK IF THAT IS THE CORRECT EQUATION NUMBER YOU ARE REFERENCING}. This type of extreme mass-ratio inspirals (EMRI) are expected to be detected by the upcoming space-based gravitational wave detector LISA. Hence, if the smaller mass component of such EMRIs contain a significant amount of net charge, then \textcolor{red}{neglecting   the extra terms generated} due to interaction of gravitational and electromagnetic radiation reactions may  lead to theoretically wrong estimation of the parameters related to these sources of gravitational waves.   
%\vspace{1cm} 
%__________________________________________________________________________________________________________________
\subsection{The significance of the term $ a^{\mu}_{5} $ :}
Now, we check the significance of the term $a^{\mu}_{5} $ with respect to the gravitational radiation reaction term $a^{\mu}_{1} $. Like the previous two terms, we first observe the ratio :
\begin{widetext}
\begin{equation} \label{5.1}
\frac{a^{\mu}_{5}}{a^{\mu}_{1}} = \frac{ \frac{2 q^{2}}{ 3 m } \xi_{1}^{2} u'^{\alpha} u'^{\rho} u'^{\sigma} (h^{R \, \mu}_{\, \, \eta} \Gamma^{\eta}_{\alpha \beta} \Gamma^{\beta}_{ \rho \sigma} + \Gamma^{\beta}_{ \rho \sigma} \Delta \Gamma^{\mu}_{\alpha \beta} + \\
\Gamma^{\mu}_{\alpha \beta} \Delta  \Gamma^{\beta}_{ \rho \sigma} + u'^{\mu} u'_{\eta} (\Gamma^{\beta}_{ \rho \sigma} \Delta \Gamma^{\eta}_{\alpha \beta} + \Gamma^{\eta}_{\alpha \beta} \Delta  \Gamma^{\beta}_{ \rho \sigma}  ))}{{\displaystyle -\xi_{1}^{2} (\delta^{\mu}_{\eta} + u^{\mu}u_{\eta}) \Delta \Gamma^{\eta}_{\alpha \beta } u'^{\alpha} u'^{\beta} }}    .
\end{equation}
\end{widetext}
So, one of the conditions for $ a^{\mu}_{5} \sim  a^{\mu}_{1}  $ is given by: 
\begin{equation} \label{5.2}
\frac{2 q^{2}}{ 3 m } \xi_{1}^{2} u'^{\alpha} u'^{\rho} u'^{\sigma} u'^{\mu} u'_{\eta} \Gamma^{\beta}_{ \rho \sigma} \Delta \Gamma^{\eta}_{\alpha \beta} \sim 
\xi_{1}^{2} u^{\mu}u_{\eta} \Delta \Gamma^{\eta}_{\alpha \beta } u'^{\alpha} u'^{\beta}    .
\end{equation}
Following the similar steps we carried out previously, we explore this condition in coordinate-wise manner, for the Reissner-Nordstrom black hole carrying constant charge\footnote{Although in many cosmological cases, the charged black holes would have charges varying continuously with time, as of now we consider black holes with constant charges.}, and after some algebra we obtain% . Similar to the steps we carried out earlier, After some algebra, we  and use the fact that  the component $ g_{rr} $ is a function of radial coordinate r only\.  Then the condition \ref{5.13} becomes :
\begin{equation}  \label{5.14}
\begin{aligned}
 \frac{ q^{2}}{ 3 m }  (u'^{r })^{2} g^{r r} \partial_{r} g_{r r }   
  \sim   \xi_{1}^{2} u'^{r}   .
\end{aligned}   
\end{equation}  
Taking $\xi \approx 1 $, and using $g_{rr} = (1- \frac{r_{s}}{r} + \frac{ r_{Q}^{2}}{r^{2}})^{-1}$, the above condition becomes\footnote{In this condition \ref{5.16}, on the LHS, we have considered the magnitude of the quantity only, neglecting the sign.}
%\begin{equation} \label{5.15}
%\frac{ q^{2}}{ 3 m }  u'^{r } g^{r r} \partial_{r} g_{r r }   
%  \sim  1 
%\end{equation}
%Here  $g_{rr} = (1- \frac{r_{s}}{r} + \frac{ r_{Q}^{2}}{r^{2}}) $ is the radial component of the metric, . % Calculating the equantity $ g^{r r} \partial_{r} g_{r r } $ for $g_{rr} = (1- \frac{r_{s}}{r} + \frac{ r_{Q}^{2}}{r^{2}}) $, we get that the above condition \ref{5.15} reduces to :
\begin{equation} \label{5.16}
\frac{ q^{2}}{ 3 m } u'^{r } \frac{  {\displaystyle \frac{1}{r} \Big(  \frac{r_{s}}{r} - \frac{2 r_{Q}^{2}}{r^{2}}  \Big) } }{ {\displaystyle   \Big(1- \frac{r_{s}}{r} + \frac{ r_{Q}^{2}}{r^{2}} \Big) } }    \sim 1    .
\end{equation}
Where, $ r_{s}$ is the Schwarzschild radius of the black hole given by $r_{s}= 2GM/c^{2} $ and $r_{Q} $ is the length scale associated with the electrical charge $Q$ of the black hole, given by $ r_{Q} = Q^{2} G / 4 \pi \epsilon_{0} c^{4} $.
For practical cases  of charged particles or compact objects moving around charged black holes,
\footnote{Here we do not specify  the sign  of the charge of  the particles. In fact, even  if  the particles's charge  are of the same sign as that of the  black hole, they may gravitationally bounded by the charged black hole if the gravitational pull due to space-time curvature overtakes the effect of electrostatic repulsion\cite{Pugliese_et_al}, \cite{Bicak_et_al}, \cite{Pugliese_et_al2}, \cite{Das_et_al}. }
 $r$ is larger than  both  these  length scales. However, the situation can be classified into two different cases : (i) $r > r_{s}, \, r_{Q} $, but yet $r \sim r_{s},  \, r_{Q} $,  and (ii) $ r >> \, r_{s}, r_{Q} $. In the former case, the quantity $  {\displaystyle \Big(  \frac{r_{s}}{r} - \frac{ 2r_{Q}^{2}}{r^{2}}  \Big) } /{ {\displaystyle   \Big(1- \frac{r_{s}}{r} + \frac{ r_{Q}^{2}}{r^{2}} \Big) } } $ is of order unity, whereas in the later case  it  would be very large. We here focus  on case (i), because if the particle moves very far away from the black hole then the effect of gravitational wave on its motion would be very little. Thus, condition (\ref{5.16}) reduces (in S.I. units) to
 % the quantity  $ \frac{  {\displaystyle \Big(  \frac{r_{s}}{r} - \frac{ r_{Q}^{2}}%{r^{2}}  \Big) } }{ {\displaystyle   \Big(1- \frac{r_{s}}{r} + \frac{ r_{Q}^{2}}{r^{2}} \Big) } } \sim 1  $. Therefore, the condition \ref{5.16} reduces to :
\begin{equation} \label{5.17}
\frac{q^{2}}{12 \pi \epsilon_{0} c^{3} m }  |u'^{r }|  \sim  r   .
\end{equation} 
This condition is similar to the one in Eqn.(\ref{4.19}), 
% and yields   identical result.% :
%\begin{equation} \label{5.18}
%| u'^{r}| \sim 10^{26} s^{-1} \, r (in \, m) 
%\end{equation}
and, hence, the cases of satisfaction of this condition would also be identical as discussed earlier. \\

Another condition of significance of $a_{5 }^{\mu} $ with respect to $a_{1 }^{\mu}$ can be given by :
\begin{equation} \label{5.19}
\frac{2 q^{2}}{ 3 m } \xi_{1}^{2}  u'^{\alpha} u'^{\rho} u'^{\sigma} \Gamma^{\beta}_{ \rho \sigma} \Delta \Gamma^{\mu}_{\alpha \beta} \sim \xi_{1}^{2}  u'^{\alpha} u'^{\beta} \Delta \Gamma^{\mu}_{\alpha \beta}   .
\end{equation}   
We expand the repeated index $ \beta $  on  both sides of the above condition, and follow similar steps using the condition coordinate-wise, as described in previous cases. Then we take the condition for the radial coordinate for the repeated-index $\beta $ and from that, cancelling the quantity $ u'^{\alpha} \Delta \Gamma^{\mu}_{\alpha r}   $ from both sides, we get  
%\begin{equation} \label{5.20}
%\begin{aligned}
%\frac{2 q^{2}}{ 3 m }  u'^{\rho} u'^{\sigma}  u'^{\alpha} ( \Gamma^{r}_{ \rho \sigma} \Delta \Gamma^{\mu}_{\alpha r}  + \Gamma^{\theta}_{ \rho \sigma} \Delta \Gamma^{\mu}_{\alpha \theta} + \Gamma^{\phi}_{ \rho %\sigma} \Delta \Gamma^{\mu}_{\alpha \phi}  +  \Gamma^{t}_{ \rho \sigma} \Delta \Gamma^{\mu}_{\alpha t} )  \\
%\sim u'^{\alpha} ( \Delta \Gamma^{\mu}_{\alpha r} u'^{r} +  \Delta \Gamma^{\mu}_{\alpha \theta} u'^{\theta} +  \Delta \Gamma^{\mu}_{\alpha \phi} u'^{\phi} +  \Delta \Gamma^{\mu}_{\alpha t} u'^{t} )
%\end{aligned}
%\end{equation}
%Considering the first term from both sides of \ref{5.20}, the condition gives :
%\begin{equation}  \label{5.21}
%\frac{2 q^{2}}{ 3 m }  u'^{\rho} u'^{\sigma}  u'^{\alpha} \Gamma^{r}_{ \rho \sigma} \Delta \Gamma^{\mu}_{\alpha r}   
%\sim    u'^{\alpha} u'^{r}  \Delta \Gamma^{\mu}_{\alpha r}    
%\end{equation}
%Cancelling the quantity $ u'^{\alpha} \Delta \Gamma^{\mu}_{\alpha r}   $ from both sides in the above condition \ref{5.21}
\begin{equation}  \label{5.22}
\frac{2 q^{2}}{ 3 m }  u'^{\rho} u'^{\sigma}  \Gamma^{r}_{ \rho \sigma}  \sim  u'^{r} .
\end{equation}  
If we expand the sum denoted by contracted repeated index $\rho $, in the LHS of above condition (\ref{5.22}), then we obtain :
\begin{eqnarray}% \label{5.23}
%\frac{2 q^{2}}{ 3 m }  u'^{r} ( u'^{\sigma}  \Gamma^{r}_{ r \sigma} ) \sim u'^{r}  
%\\
\label{5.24}
   u'^{\sigma}  \Gamma^{r}_{ r \sigma}  \sim  \frac{3m}{2q^{2}}
\end{eqnarray}
If we consider the Reissner-Nordstrom metric, then on the LHS of condition (\ref{5.24}), the only non-zero Christoffel symbol tensor component would be $   \Gamma^{r}_{ r r}   $ and hence, 
\begin{equation} \label{5.25}
u'^{r}  \Gamma^{r}_{ r r}  \sim  \frac{3m}{2q^{2}}    .
\end{equation} 
For Reissner-Nordstrom metric, the $  \Gamma^{r}_{ r r}   $ is given by :
\begin{equation} \label{5.26}
 \Gamma^{r}_{ r r}  = \frac{ {\displaystyle \frac{r_{s}}{2r} + \frac{r_{Q}^{2}}{r^{2}} }}{ {\displaystyle  r_{s} + \frac{r_{Q}^{2}}{r} - r }  }    .
\end{equation}
So, to  find out the practical cases where the condition (\ref{5.25}) holds, we need to first analyze the value of the Christoffel symbol component $\Gamma^{r}_{r r} $ around typical black holes. The value of the quantity $\frac{3m}{2q^{2}}$, which is actually $\frac{6 \pi \epsilon_{0} c^{3} m }{q^{2}} $ in S.I. units, on the RHS of condition (\ref{5.25}), is of the order of $10^{26} $ for proton. We have already calculated and used the similar quantity in condition (\ref{4.9}). There would be a limitation on the radial component of the four velocity $|u'^{r}| $, as it can not exceed the speed of light in vacuum i.e. c. Due to this limitation on $|u'^{r}| $, the order of the value of $\Gamma^{r}_{r r } $ must be higher than $10^{18} $ for satisfying the condition for the case of proton. 
\vspace{0.5cm}\\
Note  that, in the expression of $\Gamma^{r}_{rr} $ given in the Eqn.(\ref{5.26}), for any charged particle or charged compact object orbiting around a typical Reissner-Norstdorm black hole, the radial distance r must be greater than the outer horizon, which  is given by  $r_{+} = \frac{1}{2}(r_{s} + \sqrt{r_{s}^{2}-4 r_{Q}^{2} }) $. 
% WHAT DO YOU MEAN BY "THE LARGER HORIZON . 
 For stellar mass black holes, the value of $\Gamma^{r}_{rr} $ would decrease with increasing r and the same happens for supermassive black holes. But, for primordial black holes (PBHs), specially the ones having mass less than $ 10^{20} \, kg$,  exactly the opposite happens. For instance, a PBH  with  mass $\sim \, 10^{20} \, kg $  would have Schwarzschild length scale $ \sim \, 10^{-7} \, m $, and thus for any charged particle revolving around the PBH at a radial distance not exceeding $ 10^{-3.5} \, m $,  the value of $\Gamma^{r}_{rr} $ would be less than unity and this would increase with the decrease of the radial distance.  Therefore, for systems where relativistic charged particles revolve around charged PBHs of sufficiently smaller masses, created in early Universe, the condition (\ref{5.25}) could be easily satisfied resulting in significance of the term $a^{\mu}_{5 } $. In those cases even  when  the speeds of the charged particles  is about  $\, 10^{2} \, ms^{-1}$ (while it is expected to be near the speed of light  for smaller particles like proton), it would also not be difficult for the system to satisfy the condition (\ref{5.25}). 
%-------------------------------------------------------------------------------------------------
\subsection{The significance of the term $ a^{\mu }_{int \, 1} $ :}
The ratio of the interaction term $ a^{\mu }_{int \, 1} $ to the gravitational radiation reaction term $ a^{\mu}_{1}  $ is :
\begin{equation} \label{4.21}
\frac{  a^{\mu }_{int \, 1} }{ a^{\mu}_{1} } = \frac{{\displaystyle - \frac{1}{2} \frac{q}{m} F^{\mu}_{\,\, \nu} u^{\nu} u^{\alpha} u^{\beta} h^{R}_{\alpha \beta} } }{{\displaystyle -\xi_{1}^{2} (\delta^{\mu}_{\eta} + u^{\mu}u_{\eta}) \Delta \Gamma^{\eta}_{\alpha \beta } u'^{\alpha} u'^{\beta} }} \, . 
\end{equation}  
So, for $  a^{\mu }_{int \, 1} \sim a^{\mu}_{1} $, the requirement is : 
\begin{equation} \label{4.22}
 \frac{1}{2} \xi_{1}^{2} \frac{q}{m} F^{\mu}_{\,\, \nu} u^{\nu} u'^{\alpha} u'^{\beta} h^{R}_{\alpha \beta}  \sim  \xi_{1}^{2} u^{\mu}u_{\eta} \Delta \Gamma^{\eta}_{\alpha \beta } u'^{\alpha} u'^{\beta} 
\end{equation}
Using the first part of the expression of $ \Delta \Gamma^{\eta}_{\alpha \beta } $ given in the equation \ref{1.9}, in the above condition \ref{4.22}, we have :
\begin{equation} \label{4.23}
\frac{q}{m} F^{\mu}_{\,\, \eta} u^{\eta} u'^{\alpha} u'^{\beta} h^{R}_{\alpha \beta}  \sim   u^{\mu}u_{\eta} h^{R \, \eta \sigma} (\partial_{\alpha} g_{\sigma \beta } + \partial_{\beta} g_{\sigma \alpha } - \partial_{\sigma} g_{\alpha \beta} ) u'^{\alpha} u'^{\beta}  
\end{equation} 
As we did in previous cases, writing the above condition \ref{4.23} in coordinate-wise manner for $\alpha , \beta =  r $, we have :
\begin{equation} \label{4.24}
\frac{q}{m} F^{\mu}_{\,\, \eta} u^{\eta}  h^{R}_{rr} \sim u^{\mu}u_{\eta} h^{R \, \eta \sigma}  (\partial_{r} g_{\sigma r } + \partial_{r} g_{\sigma r }- \partial_{\sigma} g_{rr} )      
\end{equation}
For Reissner-Nordstrom metric, the above condition \ref{4.24} can be further simplified to :
\begin{equation} \label{4.25}
\frac{q}{m}  F^{\mu}_{\,\, \eta} u^{\eta}  h^{R}_{rr} \sim u^{\mu}u^{\eta} h^{R }_{\eta r } ( g^{rr}\partial_{r} g_{r r } )      
\end{equation}
If we expand the repeated index $\eta $ on both sides of the condition \ref{4.25} ; then again following the coordinate-wise manner and chosing the case of $ \eta = r $ only, it reduces to : 
\begin{equation} \label{4.26}
\frac{q}{m} F^{\mu}_{\,\, r} \sim u^{\mu} ( g^{rr}\partial_{r} g_{r r } )      
\end{equation}
For the Reissner-Nordstrom metric, we have already evaluated the quantity $ g^{rr}\partial_{r} g_{r r } $ in the previous sub-section with $g_{rr} =  (1- \frac{r_{s}}{r} + \frac{ r_{Q}^{2}}{r^{2}})^{-1}$ . Inserting the expression of $ g^{rr}\partial_{r} g_{r r } $ in the above condition \ref{4.26}, we obtain :
\begin{equation} \label{4.27}
\frac{q}{m} F^{\mu}_{\,\, r}  \sim u^{\mu} \frac{  {\displaystyle \frac{1}{r} \Big(  \frac{r_{s}}{r} - \frac{2 r_{Q}^{2}}{r^{2}}  \Big) } }{ {\displaystyle   \Big(1- \frac{r_{s}}{r} + \frac{ r_{Q}^{2}}{r^{2}} \Big) } }   \, .  
\end{equation}
We have already discussed about the quantity $ \frac{   \Big(  \frac{r_{s}}{r} - \frac{ 2r_{Q}^{2}}{r^{2}}  \Big)  }{  \Big(1- \frac{r_{s}}{r} + \frac{ r_{Q}^{2}}{r^{2}} \Big)  }   $ in the previous subsection, while discussing the condition of significance of the term $ a^{\mu}_{5} $. For the motion of a charged particle or a compact object near a Reissner-Nordstrom black hole, if we consider $r \sim r_{s} , \, r_{Q} $, the the above quantity is $\sim 1 $. In that case, the above condition \ref{4.27} simplifies to :
\begin{equation} \label{4.28}
\frac{q}{m}  F^{\mu}_{\,\, r}  \sim \frac{1}{r}  u^{\mu}   \,  . 
\end{equation} 
In S.I. system of units we write this condition \ref{4.28} as :
\begin{equation} \label{4.29a}
\frac{q}{4 \pi \epsilon_{0} c^{3} m}  F^{\mu}_{\,\, r}  \sim \frac{1}{r}  u^{\mu}   \,  .
\end{equation} 
For protons, the value of the quantity $ \frac{q}{4 \pi \epsilon_{0} c^{3} m} $ is $  \approx 3.19 \times 10^{-8} \, C^{-1}s $ and hence the condition \ref{4.29a} gives :
\begin{equation}  \label{4.29b}
  F^{\mu}_{\,\, r}  \sim (3.13 \times 10^{7} \, Cs^{-1}) \frac{1}{r} u^{\mu} \, .   
\end{equation} 
So, for protons or charged particles of similar category, it is expected that the above condition \ref{4.29b} is satisfied, if the charged particles move in the vicinity of a stellar-mass black hole with sufficient speed. Although, till now there is not probably any distinct observational evidence of electrically charged black holes, yet there can be external electromagnetic field, generated from any other source, near some astrophysical black holes and if this external electromagnetic field is sufficiently strong to satify the condition \ref{4.29b}, then the interaction term  $ a^{\mu }_{int \, 1} $ will be significant. 
%-----------------------------------------------------------------------------------------------------------
\subsection{The significance of the term $ a^{\mu }_{int \, 2} $ :}
From the ratio :
\begin{equation} \label{4.30}
\frac{ a^{\mu }_{int \, 2} }{ a^{\mu}_{1} } = \frac{ {\displaystyle - \frac{q}{m} (g^{\mu \nu} + u^{\mu} u^{\nu} ) h^{R}_{\nu \alpha} F^{\alpha}_{\,\, \beta } u^{\beta} }}{{\displaystyle -\xi_{1}^{2} (\delta^{\mu}_{\eta} + u^{\mu}u_{\eta}) \Delta \Gamma^{\eta}_{\alpha \beta } u'^{\alpha} u'^{\beta} }  }   \, , 
\end{equation}
we can say that one of the conditions for $  a^{\mu }_{int \, 2} \sim  a^{\mu}_{1} $ is :
\begin{equation} \label{4.31}
  \frac{q}{m}  h^{R \, \mu}_{ \alpha}  F^{\alpha}_{\,\, \beta } u^{\beta} \sim \Delta \Gamma^{\mu}_{\alpha \beta } u^{\alpha} u^{\beta}
\end{equation}
Using the first part of the expression of $ \Delta \Gamma^{\mu}_{\alpha \beta } $ given in the equation \ref{1.9}, in the above condition \ref{4.31}, we have :
\begin{equation} \label{4.32}
\frac{q}{m}  h^{R \, \mu \alpha} F_{\alpha \beta } u^{\beta} \sim  h^{R \, \mu \sigma} (\partial_{\alpha} g_{\sigma \beta } + \partial_{\beta} g_{\sigma \alpha } - \partial_{\sigma} g_{\alpha \beta} )  u^{\alpha} u^{\beta} \, . 
\end{equation} 
Now similar to the previous cases, considering the above condition \ref{4.32} in coordinate-wise manner for $ \alpha , \beta = r  $, we get :
\begin{equation} \label{4.33}
\frac{q}{m}  h^{R \, \mu r} F_{ r r } \sim  h^{R \, \mu \sigma} (\partial_{r} g_{\sigma r } + \partial_{r} g_{\sigma r} - \partial_{\sigma} g_{r r} )  u^{r}  \,   . 
\end{equation} 
For diagonal metrics like Reissner-Nordstrom metric, the above condition \ref{4.33} will be further simplified to  :
\begin{equation} \label{4.34}
\frac{q}{m}  h^{R \, \mu r} F_{ r r } \sim  h^{R \, \mu r}  u^{r}  \partial_{r} g_{r r }   
\end{equation} 
or, 
\begin{equation} \label{4.35}
\frac{q}{m}  F_{ r r } \sim   u^{r}  \partial_{r} g_{r r }    \,   . 
\end{equation}
So, for Reissner-Nordstrom metric, the above condition \ref{4.35} becomes :
\begin{equation} \label{4.36}
\frac{q}{m}  F_{ r r } \sim   u^{r} \frac{\frac{1}{r} \Big( \frac{r_{s}}{r} - \frac{ 2r_{Q}^{2}}{r^{2}}  \Big)}{\Big(1 - \frac{r_{s}}{r} + \frac{r_{Q}^{2}}{r^{2}} \Big)^{2}}    \,  . 
\end{equation}
For charged particles or charged compact objects moving near the black hole in such a way that $ r \sim r_{s} ,  \, r_{Q} $, the quantity $ \Big( \frac{r_{s}}{r} - \frac{ 2r_{Q}^{2}}{r^{2}}  \Big) / \Big(1 - \frac{r_{s}}{r} + \frac{r_{Q}^{2}}{r^{2}} \Big)^{2} $ in the R.H.S. of the above condition \ref{4.36} is $ \sim 1$. Then in that case, the above condition simplifies to :
\begin{equation} \label{4.37}
\frac{q}{m}  F_{ r r } \sim   \frac{u^{r}}{r} \, ,  
\end{equation} 
or in S.I. system of units, it can be written as :
\begin{equation} \label{4.38}
\frac{q}{ 4 \pi \epsilon_{0} c^{3} m}  F_{ r r } \sim   \frac{u^{r}}{r}  \,   . 
\end{equation}
Similarly as before, substituting the value of the quantity $ \frac{q}{ 4 \pi \epsilon_{0} c^{3} m} $ for proton, we write the condition \ref{4.38} as :
\begin{equation} \label{4.39}
   F_{ r r } \sim  (3.13 \times 10^{7} \, Cs^{-1}) \frac{u^{r}}{r}   \,  , 
\end{equation} 
which is quite similar to the previous case of the condition \ref{4.29b}. So, for the same cases, as discussed in the prvious subsection, this condition too would be satisfied. 
%_________________________________________________________________________________________________
\section{Conclusion and Discussion : }
In this work we have shown that coexistence of metric perturbations and electromagnetic self-force can lead to an effect, in the motion of charged particles in curved space-time, which does not exist when any one among these two is absent. In most of the physical situations, the metric perturbation is the gravitational radiation emitted from the charged particle itself, which causes the gravitational self-force.
The perturbative terms of the electromagnetic self-force, which we have derived, are different from the interaction terms of electromagnetic and gravitational self-forces, given in the work of P. Zimmerman and E. Poisson \cite{Zimmerman_et_al}. \\
 We have analyzed different conditions for which these perturbative terms generated from the electromagnetic self-force due to its perturbation by the gravitational radiation would be significant in comparison with the gravitational self-force. We have also analyzed the conditions of significance of the interaction terms of electromagnetic and gravitational self-forces in comparison with the gravitational self-force.\\ 
  It is interesting to find that there are astrophysical phenomena and cosmological cases where these perturbative terms can play a significant role. The physical interpretation of these perturbations to the electromagnetic self-force by the gravitational radiation can be understood as the fact that, when electromagnetic self-force is acting in curved space-time, where there is gravitational wave emission from the system, then the electromagnetic wave produced due to the motion of the charged particle has to traverse through the ripples in the curved space-time due to the gravitational radiation. However, in the absence of the gravitational radiation, the electromagnetic wave propagates through the curved space-time but it does not face ripples in space-time. 
It is this difference with the case where gravitational radiation is present, that manifests in the form of these perturbative terms generated in the equation of motion of the charged particle. It is important to note that not taking into account these perturbative terms in the specified astrophysical or cosmological phenomena involving relativistic charged particles or compact objects, and considering only the gravitational radiation reaction can
 %ing whether they are negligible or not, will 
lead to incorrect estimation of their motions. \\
In this way, we have also demonstrated that it can be misleading if we estimate or compare magnitude of terms in the equation-of-motion of the charged particle, only in terms of $q $ and $m $. The other physical quantities present in the terms also matter. They can play a significant role in determining the overall order of certain term. We have not dicussed the case where the source of gravitational radiation perturbing the system is external. But, from the study in this work we can say, that for a suitable external source of gravitational radiation too, the perturbative terms of the electromagnetic self-force can be significant in comparison with the gravitational self-force.      
\section{Acknowledgement}
Arnab Sarkar thanks S. N. Bose National Centre for Basic Sciences, Kolkata 700106, under Department of Science and Technology, Govt. of India, for funding through institute-fellowship. Amna Ali is thankful to UGC for providing financial support under the scheme Dr. D.S. Kothari postdoctoral fellowship. Arnab Sarkar is grateful for the valuable advice and comments of Dr. Adam Pound, Royal Society University Research Fellow within Mathematical Sciences at the University of Southampton, the UK. Dr. Adam Pound's suggestions have helped greatly to improve this work and it is a 
privilege to get his opinions.    \\
Also, Arnab Sarkar wants to thank International Centre for Theoretical Sciences (ICTS) TIFR, Bengaluru 560089, India, as the motivation for this work came after attending a lecture series on `Self-force and radiation reaction in general relativity' by Dr. Adam Pound, as a part of the programme -`` Summer School on Gravitational Wave Astronomy 2019 " (Code: ICTS/gws2019/07).
%_____________________________________________________________________________________________________________
%_____________________________________________________________________________________________________________
\section{APPENDIX-1 : Orthogonality of the radiation reaction terms with four-velocity :}
\subsection{The orthogonality properties of different terms in $a^{\mu} $ :}
Now, we test the orthogonality of the radiation reaction term $a^{\mu} $. We just write the expression of $a^{\mu} $ from Eqn.(\ref{1.3}) in the following way :
\begin{equation} \label{2.1}
\begin{aligned}
a^{\mu}= \frac{D u^{\mu}}{d \tau} -  \frac{q}{m}  F^{\mu}_{\,\, \nu} u^{\nu} + \frac{2q^{2}}{3m} \Big( \frac{D^{2} u^{\mu}}{d\tau^{2}} + u^{\mu} u_{\nu} \frac{D^{2} u^{\nu}}{d \tau^{2}} \Big)   + 
\\  
\frac{q^{2}}{3m} (R^{\mu}_{\lambda} u^{\lambda} + R^{\nu}_{\lambda} u^{\lambda} u^{\mu} u_{\nu} ) +    \frac{2q^{2}}{m} f^{\mu \nu}_{Tail} u_{\nu}   .
\end{aligned} 
\end{equation}
It is already known that the part $\frac{D u^{\mu}}{d \tau} $ is perpendicular to the four-velocity $u^{\mu} $ \cite{Barack} i.e. 
\begin{equation} \label{2.2}
\frac{D u^{\mu}}{d \tau}  u_{\mu} = \Big( \frac{d^{2} x^{\mu}}{d \tau^{2}} + \Gamma^{\mu}_{\nu \rho} \frac{d x^{\nu}}{d \tau} \frac{d x^{\rho}}{d \tau}  \Big) u_{\mu} = 0      .
\end{equation}
Next, we test the orthogonality with the electromagnetic radiation reaction term $ \frac{2q^{2}}{3m} ( g^{\mu}_{\nu} + u^{\mu} u_{\nu} )\frac{D^{2} u^{\nu}}{d \tau^{2}}   $, using the relation $u^{\alpha}u_{\alpha}=  -1 $ :
\begin{equation}  \label{2.3}
\begin{aligned}
{\displaystyle \Big( g^{\mu}_{\nu} + u^{\mu} u_{\nu} \Big) \frac{D^{2} u^{\nu}}{d \tau^{2}}  u_{\mu} =  \Big( g^{\mu}_{\nu}  u_{\mu} + u^{\mu} u_{\mu} u_{\nu}  \Big) \frac{D^{2} u^{\nu}}{d \tau^{2}}    } 
\\
{\displaystyle   = (u_{\nu} + (-1) u_{\nu} ) \frac{D^{2} u^{\nu}}{d \tau^{2}} = 0    }   .
\end{aligned}
\end{equation}
The orthogonality of the Lorentz force term $\frac{q}{m}F^{\mu}_{\,\, \nu} u^{\nu} $ results from the antisymmetry of the field strength tensor under the exchange of the space-time indices.
Then, we test the orthogonality of the tail term $ \frac{2q^{2}}{m} f^{\mu \nu}_{Tail} u_{\nu} $. The $ f^{\mu \nu}_{Tail} $ in the tail term is the `tail integral' given by \cite{Poisson, Tursunov_et_al}: 
\begin{equation} \label{2.6}
f^{\mu \nu}_{Tail} = \int_{- \infty}^{\tau - 0^{+}} D^{[ \mu } G^{\nu ]}_{+\lambda''} (z(\tau) , z(\tau'')) u^{\lambda''} d\tau''  .
\end{equation}
Where $ G^{\mu}_{+ \lambda} $ is the retarded Green's function associated with the vector potential of the electromagnetic field. 
Hence, contracting the term $ f^{\mu \nu}_{Tail} u_{\nu} $ with $ u_{\mu}$ we have:
%\vspace{1cm}
\begin{equation} \label{2.7}
\begin{aligned}
f^{\mu \nu}_{Tail} u_{\nu} u_{\mu} =
  u_{\nu} u_{\mu}\int_{- \infty}^{\tau - 0^{+}} D^{[ \mu } G^{\nu ]}_{+\lambda''} (z(\tau) , z(\tau'')) u^{\lambda''} d\tau''   
  \\
=  u_{\nu} u_{\mu}\int_{- \infty}^{\tau - 0^{+}} ( D^{ \mu } G^{\nu }_{+\lambda''} - D^{ \nu } G^{\mu }_{+\lambda''} )  (z(\tau) , z(\tau'')) u^{\lambda''} d\tau''  
\\
=   u_{\nu} u_{\mu}\int_{- \infty}^{\tau - 0^{+}}  D^{ \mu } G^{\nu }_{+\lambda''}  (z(\tau) , z(\tau'')) u^{\lambda''} d\tau''
\\
  - u_{\nu} u_{\mu}\int_{- \infty}^{\tau - 0^{+}}  D^{ \nu } G^{\mu }_{+\lambda''}  (z(\tau) , z(\tau'')) u^{\lambda''} d\tau''  .
\end{aligned}    
\end{equation}
In this case also, in the two parts on the RHS of the above Eqn.(\ref{2.7}), the indices $\mu $ and $\nu $ are repeated and in the similar way, as can be done for the previous case of Lorentz force, in this case also we interchange the indices ($\mu   \leftrightarrow \nu $) for the first term on the RHS of Eqn.(\ref{2.7}) and obtain :
\begin{equation} \label{2.8}
\begin{aligned}
f^{\mu \nu}_{Tail} u_{\nu} u_{\mu}
 =   u_{\nu} u_{\mu}\int_{- \infty}^{\tau - 0^{+}}  D^{ \mu } G^{\nu }_{+\lambda''}  (z(\tau) , z(\tau'')) u^{\lambda''} d\tau''
 \\
  - u_{\nu} u_{\mu}\int_{- \infty}^{\tau - 0^{+}}  D^{ \nu } G^{\mu }_{+\lambda''}  (z(\tau) , z(\tau'')) u^{\lambda''} d\tau'' 
\\
=  u_{\mu} u_{\nu}\int_{- \infty}^{\tau - 0^{+}}  D^{ \nu } G^{\mu }_{+\lambda''}  (z(\tau) , z(\tau'')) u^{\lambda''} d\tau''
\\
  - u_{\nu} u_{\mu}\int_{- \infty}^{\tau - 0^{+}}  D^{ \nu } G^{\mu }_{+\lambda''}  (z(\tau) , z(\tau'')) u^{\lambda''} d\tau''
  = 0   .
 \end{aligned}
\end{equation} 
Hence, the overall radiation reaction term $a^{\mu} $ is orthogonal to the four-velocity $ u_{\mu}$ : $a^{\mu} u_{\mu} = 0$.
In the next sub-section we shall show the utility of this orthogonality property.
%\vspace{0.3cm} 
%_____________________________________________________________________________________________________
\subsection{Utility of the Orthogonality Property of the reaction $ a^{\mu}$ with the four-velocity :}
As we have the orthogonality property of the overall radiation reaction, we can use it to have constraints or relations between different coefficients and terms present in the reaction. For doing this, we contract the radiation reaction $a^{\mu} $ with unperturbed four-velocity $ u_{\mu}$ in the Eqn.(\ref{1.7}). Thus we get :
%\vspace{0.2cm}
\begin{widetext}
\begin{equation} \label{3.1}
\begin{aligned}
\left\lbrace    \frac{d^{2} \tau'}{d \tau^{2}} \frac{d \tau }{d \tau' }\frac{d x^{\mu}}{d \tau} +  (- \Delta\Gamma^{\mu}_{\nu \rho}) \frac{ d x^{\nu}}{d \tau}\frac{ d x^{\rho}}{d \tau}  \right\rbrace u_{\mu}  +  \Big( \frac{d \tau'}{d \tau} \Big)  \frac{q}{m} \big( F'^{\mu \nu} - \frac{d \tau}{d \tau'} F^{\mu \nu} \big) u_{\nu} u_{\mu} + 
 % \Big( \frac{d \tau'}{d \tau} \Big)^{2} \frac{q^{2}}{3 m}  ((R'^{\mu}_{\lambda} u'^{\lambda} + R'^{\nu}_{\lambda} u'^{\lambda} u'^{\mu} u'_{\nu} ) -  \Big(\frac{d \tau'}{d \tau} \Big)^{-2}(R^{\mu}_{\lambda} u^{\lambda} + R^{\nu}_{\lambda} u^{\lambda} u^{\mu} u_{\nu} )) +
 \\
 \frac{2 q^{2}}{m} \Big(\frac{d \tau'}{d \tau} \Big) \Big(f'^{\mu \nu}_{Tail} -  \frac{d \tau}{d \tau'} f^{\mu \nu}_{Tail} \Big) u_{\nu} u_{\mu} +
\frac{2q^{2}}{3m} \Big(\frac{d \tau'}{d \tau} \Big)^{2} \frac{d^{3} x^{\eta}}{d \tau'^{3}} 
\left\lbrace \Big( g'^{\mu}_{\eta} +   \frac{d x^{\mu}}{ d \tau'}  \frac{d x_{\eta}}{ d \tau'} \Big) u_{\mu}   - \frac{d \tau'}{d \tau} (0)  \right\rbrace    +
\\ 
\frac{2q^{2}}{3m} \Big(\frac{d \tau'}{d \tau} \Big)^{2}  \frac{ d x^{\alpha}}{d \tau'}\frac{ d x^{\beta}}{d \tau'}\frac{ d x^{\gamma}}{d \tau'} 
\left\lbrace \Big( g'^{\mu}_{\eta} +   \frac{d x^{\mu}}{ d \tau'}  \frac{d x_{\eta}}{ d \tau'} \Big)  \partial_{\gamma} \Gamma'^{\eta}_{\alpha \beta} u_{\mu} -  \frac{d \tau'}{d \tau}  (0)  \right\rbrace   +
\\
\frac{2q^{2}}{3m} \Big(\frac{d \tau'}{d \tau} \Big)^{2}  \frac{ d x^{\alpha}}{d \tau'}\frac{ d^{2} x^{\beta}}{d \tau'^{2}}
\left\lbrace \Big( g'^{\mu}_{\eta} +   \frac{d x^{\mu}}{ d \tau'}  \frac{d x_{\eta}}{ d \tau'} \Big)  3 \Gamma'^{\eta}_{\alpha \beta}  u_{\mu}-  \frac{d \tau'}{d \tau} (0)  \right\rbrace   +
\\
\frac{2q^{2}}{3m} \Big(\frac{d \tau'}{d \tau} \Big)^{2} \frac{ d x^{\alpha}}{d \tau'}\frac{ d x^{\rho}}{d \tau'}\frac{ d x^{\sigma}}{d \tau'}
  \left\lbrace    \Big( g'^{\mu}_{\eta} +   \frac{d x^{\mu}}{ d \tau'}  \frac{d x_{\eta}}{ d \tau'} \Big)   \Gamma'^{\eta}_{\alpha \beta} \Gamma'^{\beta}_{\rho \sigma} u_{\mu} -  \frac{d \tau'}{d \tau}(0)  \right\rbrace  -
  \\
  \frac{2q^{2}}{3m}  \Big( g'^{\mu}_{\eta} +   \frac{d x^{\mu}}{ d \tau'}  \frac{d x_{\eta}}{ d \tau'} \Big)         \left\lbrace  3 \Gamma^{\eta}_{\alpha \beta}  \frac{d \tau'}{d \tau} \frac{d^{2} \tau'}{d \tau^{2}} \Big(  \frac{d x^{\alpha}}{ d \tau'}\frac{d x^{\beta}}{ d \tau'} \Big) +   \frac{d^{3} \tau'}{d \tau^{3}} \frac{d \tau'}{d \tau} \frac{d x^{\eta}}{d \tau'}  +  3   \frac{d \tau'}{d \tau} \frac{d^{2} \tau'}{d \tau^{2}} \frac{d^{2} x^{\eta}}{d \tau'^{2}}  \right\rbrace u_{\mu}
  = a^{\mu}u_{\mu}     .
\end{aligned}
\end{equation} 
We have already shown that $F^{\mu \nu} u_{\nu} u_{\mu} = 0  $ and $ f^{\mu \nu}_{Tail} u_{\mu} u_{\nu} =  0 $ in the previous sub-section. The similar results are also valid for the perturbed external lorentz force and the perturbed tail term : $F'^{\mu \nu} u_{\nu} u_{\mu} = 0  $ and $ f'^{\mu \nu}_{Tail} u_{\mu} u_{\nu} =  0 $. Using those results and the fact that $ \Big( g^{\mu}_{\eta} + \frac{d x^{\mu}}{ d \tau}  \frac{d x_{\eta}}{ d \tau} \Big)u_{\mu} = 0  $, in the above Eqn.(\ref{3.1}), we can write it in the following way :   
\begin{equation} \label{3.2}
\begin{aligned}
\left\lbrace    \frac{d^{2} \tau'}{d \tau^{2}} \frac{d \tau }{d \tau' } (-1)+  (- \Delta\Gamma^{\mu}_{\nu \rho}) u_{\mu} \frac{ d x^{\nu}}{d \tau}\frac{ d x^{\rho}}{d \tau}  \right\rbrace   +  
\frac{2q^{2}}{3m} \Big(\frac{d \tau'}{d \tau} \Big)^{2} \frac{d^{3} x^{\eta}}{d \tau'^{3}} 
\left\lbrace \Big( g'^{\mu}_{\eta} +   \frac{d x^{\mu}}{ d \tau'}  \frac{d x_{\eta}}{ d \tau'} \Big) u_{\mu}    \right\rbrace    + 
\\
\frac{2q^{2}}{3m} \Big(\frac{d \tau'}{d \tau} \Big)^{2}  \frac{ d x^{\alpha}}{d \tau'}\frac{ d x^{\beta}}{d \tau'}\frac{ d x^{\gamma}}{d \tau'} 
\left\lbrace \Big( g'^{\mu}_{\eta} +   \frac{d x^{\mu}}{ d \tau'}  \frac{d x_{\eta}}{ d \tau'} \Big)  \partial_{\gamma} \Gamma'^{\eta}_{\alpha \beta} u_{\mu}  \right\rbrace   +
\\
\frac{2q^{2}}{3m} \Big(\frac{d \tau'}{d \tau} \Big)^{2}  \frac{ d x^{\alpha}}{d \tau'}\frac{ d^{2} x^{\beta}}{d \tau'^{2}}
\left\lbrace \Big( g'^{\mu}_{\eta} +   \frac{d x^{\mu}}{ d \tau'}  \frac{d x_{\eta}}{ d \tau'} \Big)  3 \Gamma'^{\eta}_{\alpha \beta}  u_{\mu} \right\rbrace   +
\\
\frac{2q^{2}}{3m} \Big(\frac{d \tau'}{d \tau} \Big)^{2} \frac{ d x^{\alpha}}{d \tau'}\frac{ d x^{\rho}}{d \tau'}\frac{ d x^{\sigma}}{d \tau'}
  \left\lbrace    \Big( g'^{\mu}_{\eta} +   \frac{d x^{\mu}}{ d \tau'}  \frac{d x_{\eta}}{ d \tau'} \Big)   \Gamma'^{\eta}_{\alpha \beta} \Gamma'^{\beta}_{\rho \sigma} u_{\mu}   \right\rbrace  -
  \\
  \frac{2q^{2}}{3m}  \Big( g'^{\mu}_{\eta} +   \frac{d x^{\mu}}{ d \tau'}  \frac{d x_{\eta}}{ d \tau'} \Big)         \left\lbrace  3 \Gamma^{\eta}_{\alpha \beta}  \frac{d \tau'}{d \tau} \frac{d^{2} \tau'}{d \tau^{2}} \Big(  \frac{d x^{\alpha}}{ d \tau'}\frac{d x^{\beta}}{ d \tau'} \Big) +   \frac{d^{3} \tau'}{d \tau^{3}} \frac{d \tau'}{d \tau} \frac{d x^{\eta}}{d \tau'}  +  3   \frac{d \tau'}{d \tau} \frac{d^{2} \tau'}{d \tau^{2}} \frac{d^{2} x^{\eta}}{d \tau'^{2}}  \right\rbrace u_{\mu}
  = 0    .
\end{aligned}
\end{equation}
Now, we simplify the expression $ \Big( g'^{\mu}_{\eta} +   \frac{d x^{\mu}}{ d \tau'}  \frac{d x_{\eta}}{ d \tau'} \Big) u_{\mu } $ :
\begin{equation} \label{3.3}
\begin{aligned}
\Big( g'^{\mu}_{\eta} +   \frac{d x^{\mu}}{ d \tau'}  \frac{d x_{\eta}}{ d \tau'} \Big) u_{\mu} 
=  (g^{\mu}_{\eta} + h^{R \, \mu}_{\eta})u_{\mu} +  \Big( \frac{d \tau}{ d \tau'} \Big)^{2} \frac{d x^{\mu}}{ d \tau}  \frac{d x_{\eta}}{ d \tau} u_{\mu}  
\\
= (g^{\mu}_{\eta}u_{\mu} + h^{R \, \mu}_{\eta}u_{\mu}) +  \Big( \frac{d \tau}{ d \tau'} \Big)^{2}  u_{\eta} (u^{\mu}u_{\mu} ) 
\\
 = (u_{\eta} + h^{R \, \mu}_{\eta}u_{\mu}  ) +  \Big(\frac{d \tau}{ d \tau'} \Big)^{2} u_{\eta} (-1)  
= u_{\eta} \Big(  1 - \Big(\frac{d \tau}{ d \tau'} \Big)^{2}  \Big) + h^{R \, \mu}_{\eta}u_{\mu}    .
\end{aligned}  
\end{equation}
Using this above expression and simplifying the Eqn(\ref{3.2}), we obtain :
%\vspace{2cm}\\
\begin{equation} \label{3.4}
\begin{aligned}
 -\left\lbrace  \frac{d^{2} \tau'}{d \tau^{2}} \frac{d \tau }{d \tau' } +   \Delta\Gamma^{\mu}_{\nu \rho} u_{\mu} u^{\nu} u^{\rho}  \right\rbrace =
-\left\lbrace  u_{\eta} \Big(  1 - \Big(\frac{d \tau}{ d \tau'} \Big)^{2}  \Big) + h^{R \, \mu}_{\eta}u_{\mu} \right\rbrace  \frac{2 q^{2}}{3 m }
\\
 \Big[ \Big(\frac{d \tau'}{d \tau} \Big)^{2} \Big(  \frac{d^{3} x^{\eta}}{d \tau'^{3}}  +
 u'^{\alpha} u'^{\beta} u'^{\gamma} \partial_{\gamma} \Gamma'^{\eta}_{\alpha \beta} + 3 u'^{\alpha} \frac{d u'^{\beta}}{d \tau'} \Gamma'^{\eta}_{\alpha \beta} + u'^{\alpha} u'^{\rho} u'^{\sigma} \Gamma'^{\eta}_{\alpha \beta} \Gamma'^{\beta}_{\rho \sigma} \Big)      
 \\
 -\Big(  3 \Gamma^{\eta}_{\alpha \beta}  \frac{d \tau'}{d \tau} \frac{d^{2} \tau'}{d \tau^{2}} \Big(  \frac{d x^{\alpha}}{ d \tau'}\frac{d x^{\beta}}{ d \tau'} \Big) +   \frac{d^{3} \tau'}{d \tau^{3}} \frac{d \tau'}{d \tau} \frac{d x^{\eta}}{d \tau'}  +  3   \frac{d \tau'}{d \tau} \frac{d^{2} \tau'}{d \tau^{2}} \frac{d^{2} x^{\eta}}{d \tau'^{2}}  \Big)       
\Big]  .
\end{aligned}
\end{equation}
\end{widetext}
\vspace{0.5cm}
Now, using the above Eqn.(\ref{3.4}) we can substitute the terms $ \Big(  3 \Gamma^{\eta}_{\alpha \beta}  \frac{d \tau'}{d \tau} \frac{d^{2} \tau'}{d \tau^{2}} \Big(  \frac{d x^{\alpha}}{ d \tau'}\frac{d x^{\beta}}{ d \tau'} \Big) +   \frac{d^{3} \tau'}{d \tau^{3}} \frac{d \tau'}{d \tau} \frac{d x^{\eta}}{d \tau'}  +  3   \frac{d \tau'}{d \tau} \frac{d^{2} \tau'}{d \tau^{2}} \frac{d^{2} x^{\eta}}{d \tau'^{2}}  \Big)   $ in the Eqn.(\ref{1.8}) in terms of $g_{\mu \nu}, \, h_{\mu\nu}, \Gamma^{\mu}_{\nu \rho} , \Delta\Gamma^{\mu}_{\nu \rho}, u_{\mu} $ etc. 
%___________________________________________________________________________________________________________
\section{APPENDIX-2 : The reason for neglecting the term containing the Ricci-tensor : }
%The referee has said that the term containing the Ricci-tensor, which arises due to the interaction with the surrounding matter, is not zero due to the presence of the electromagnetic field. However, we have not said that it vanishes, we have only neglected the term. Here, we give the reason that why we neglect it.
As there is a background electromagnetic field in this case due to the external electromagnetic field and also due to the electromagnetic field emitted by the charged particle, so the term$ \frac{q^{2}}{3m} ( R^{\mu }_{\lambda} u^{\lambda} + R^{\nu}_{\lambda} u^{\lambda} u^{\mu} u_{\nu}  ) $ is in general non-zero. But, we can have an idea of its order by some analysis. We show this here.\\  
 As the electromagnetic field is a kind of radiation, with EoS-parameter w= $1/3$, the Ricci-scalar due to it vanishes and this can be shown simply from the Einstein's equation viz. the Einstein-Maxwells' equation here.\\ 
We start from the Einstein-Maxwell's equation or the Einstein's equation with electromagnetic stress-energy tensor as the source : 
\begin{equation} \label{A2.1}
R_{\mu \nu} - \frac{1}{2} g_{\mu \nu} R = \frac{8 \pi G}{c^{4} \mu_{o}} (  F_{\mu}^{\,\, \eta }  F_{\nu\eta} - \frac{1}{4} g_{\mu \nu} F^{2} )   \, ,
\end{equation}
where $\mu_{0} $ is the permeability in vacuum and $ F_{\mu \nu}$ is the electromagnetic field tensor.
Contracting both sides of the above equation \ref{A2.1} with $g^{\mu \nu} $, we get : 
\begin{eqnarray} 
\begin{aligned}
R_{\mu \nu}g^{\mu \nu} - \frac{1}{2} g_{\mu \nu}g^{\mu \nu} R = 
\\
\frac{8 \pi G}{c^{4} \mu_{o}} (  F_{\mu}^{\,\, \eta }  F_{\nu\eta} g^{\mu \nu}- \frac{1}{4} g_{\mu \nu}g^{\mu \nu} F^{2} )  \, , \\
\Rightarrow  R - \frac{1}{2}(4) R = \frac{8 \pi G}{  c^{4} \mu_{o}} ( F_{\mu}^{\,\, \eta }  F^{\mu}_{\,\, \eta} - \frac{1}{4}(4) F^{2} ) \, \\
\Rightarrow - R =  \frac{8 \pi G}{c^{4} \mu_{o}} (F^{2} - F^{2} ) = 0  \label{A2.2}
\end{aligned}
\end{eqnarray} 
So, we have shown that for electromagnetic field the Ricci-scalar R is zero (0). Now, we substitute R = 0 in the equation \ref{A2.1} and get :
\begin{equation} 
R_{\mu \nu}  = \frac{8 \pi G}{c^{4} \mu_{o}} (  F_{\mu}^{\,\, \eta } F_{\nu\eta} - \frac{1}{4} g_{\mu \nu} F^{2} )   \, , 
\end{equation}
Or, in mixed-tensorial form, 
\begin{equation}
R^{\mu}_{\nu} =  \frac{8 \pi G}{ c^{4} \mu_{o}}  (F^{\mu \eta } F_{\nu\eta} - \frac{1}{4} g^{\mu}_{\nu} F^{2} ) \,  . \label{A2.3}
\end{equation} 
The term containing Ricci-tensors, appearing in the equation of motion of a charged particle in curved space-time, is 
$\frac{q^{2}}{3m}( R^{\mu }_{\lambda} u^{\lambda} + R^{\nu}_{\lambda} u^{\lambda} u^{\mu} u_{\nu} ) $. We first evaluate the first part $ R^{\mu }_{\lambda} u^{\lambda} $ : 
\begin{equation} \label{A2.4}
\begin{aligned}
R^{\mu }_{\lambda} u^{\lambda} = \frac{8 \pi G}{c^{4} \mu_{o}}  (F^{\mu \eta } F_{\lambda\eta} u^{\lambda} - \frac{1}{4} g^{\mu}_{\lambda} u^{\lambda} F^{2} ) 
\\
= \frac{8 \pi G}{ c^{4} \mu_{o}}  (F^{\mu \eta } F_{\lambda\eta} u^{\lambda} - \frac{1}{4} u^{\mu} F^{2} ) \, . 
\end{aligned} 
\end{equation}   
Then, we evaluate the second part : 
\begin{eqnarray} \label{A2.5}
R^{\nu}_{\lambda} u^{\lambda} u^{\mu} u_{\nu}  = \frac{8 \pi G}{ c^{4} \mu_{o}}  (F^{\nu \eta } F_{\lambda\eta} u^{\lambda} - \frac{1}{4} g^{\nu}_{\lambda} u^{\lambda} F^{2} )  u^{\mu} u_{\nu}  \\
%\Rightarrow R^{\nu}_{\lambda} u^{\lambda} u^{\mu} u_{\nu} =  \frac{8 \pi G}{ c^{4} \mu_{o}}  (F^{\nu \eta } F_{\lambda\eta} u^{\lambda}  u^{\mu} u_{\nu}- \frac{1}{4}  u^{\mu} u_{\nu} u^{\nu} F^{2} )  \\
\Rightarrow R^{\nu}_{\lambda} u^{\lambda} u^{\mu} u_{\nu}  =  \frac{8 \pi G}{  c^{4} \mu_{o}}  (F^{\nu \eta } F_{\lambda\eta} u^{\lambda}  u^{\mu} u_{\nu} + \frac{1}{4}  u^{\mu} F^{2}  )  \, . 
\end{eqnarray}
Hence, 
\begin{equation} \label{A2.6}
\begin{aligned}
R^{\mu }_{\lambda} u^{\lambda} + R^{\nu}_{\lambda} u^{\lambda} u^{\mu} u_{\nu}  =   
\\
\frac{8 \pi G}{  c^{4} \mu_{o}}  (F^{\mu \eta } F_{\lambda\eta} u^{\lambda} - \frac{1}{4} u^{\mu} F^{2}  + F^{\nu \eta } F_{\lambda\eta} u^{\lambda}  u^{\mu} u_{\nu} + \frac{1}{4}  u^{\mu} F^{2}   )   \\
= \frac{8 \pi G}{ c^{4} \mu_{o}}  (F^{\mu \eta }  + F^{\nu \eta }  u^{\mu} u_{\nu}  ) F_{\lambda\eta} u^{\lambda}  
=  \frac{8 \pi G}{ c^{4} \mu_{o}} ( \delta^{\mu}_{\nu} + u^{\mu} u_{\nu} ) F^{\nu \eta } F_{\lambda\eta} u^{\lambda}  
\end{aligned}
\end{equation}
Hence, 
\begin{equation} \label{A2.7}
\begin{aligned}
\frac{q^{2}}{12 \pi \epsilon_{0} c^{3} m } ( R^{\mu }_{\lambda} u^{\lambda} + R^{\nu}_{\lambda} u^{\lambda} u^{\mu} u_{\nu}  )
\\
= \frac{q^{2}}{12 \pi \epsilon_{0} c^{3} m} \frac{8 \pi G}{ c^{4} \mu_{o}} ( \delta^{\mu}_{\nu} + u^{\mu} u_{\nu} ) F^{\nu \eta } F_{\lambda\eta} u^{\lambda}  \, , 
\end{aligned}
\end{equation}
where, we have written the coefficient $q^{2}/3m $ , as written in general calculations of our manuscript, in S.I. units as $\frac{q^{2}}{12 \pi \epsilon_{0} c^{3} m} $ ; $ \epsilon_{0}$ being the permittivity of vacuum. It is to be noted that the term $q F_{\lambda\eta} u^{\lambda} $ within the R.H.S.(right hand side) of the equation \ref{A2.7} is just the Lorentz-force due to the electromagnetic field $ F_{\lambda \eta}$. Among the rest of the terms $  ( \delta^{\mu}_{\nu} + u^{\mu} u_{\nu} ) $ has the order $ \sim 1 $. Now, the rest coefficient is :
\begin{equation} \label{A2.8}
 \frac{8 \pi G }{  12 \pi \epsilon_{0} \mu_{o} c^{7} } \frac{q}{m}F^{\nu \eta}  = \frac{2 G }{  3 c^{5} } \frac{q}{m} F^{\nu \eta} \,  , 
\end{equation}       
as we know that $\epsilon_{0} \mu_{0} = c^{-2} $. 
Evaluating the numerical value of the quantity $ \frac{2 G }{  3 c^{5} }  $, we get : 
\begin{equation}
\frac{2 G }{  3 c^{5} } \approx  1.83 \times 10^{-53} \, kg^{-1} m^{-2} s^{3}   \,  . 
\end{equation}
Therefore, there can hardly be any astrophysical and cosmological cases, which we have discussed in our work, having charge-to-mass ratio so high, that it can compensate the very much small factor $ 10^{-53} $ . On the other hand it is hardly possible to find any astrophysical or cosmological case where the external electric or magnetic field is so huge that the electromagnetic-field can compensate the factor $  10^{-53}  $. \\
Generally, the astrophysical compact objects like neutron stars, white dwarfs and black holes can have very high magnetic fields $\sim 10^{17} \, G $ or $10^{13} \, T $, associated with them. On the other hand, we have already stated that theoretically maximum-possible value of electric fields, that can be produced by charged neutron stars near them, is $\sim 10^{21} V/m$. So, it is quite clear that even these very high magnetic fields and electric fields are also insufficient to make this term $\frac{q^{2}}{3m}( R^{\mu }_{\lambda} u^{\lambda} + R^{\nu}_{\lambda} u^{\lambda} u^{\mu} u_{\nu} ) $ significant, overcoming the extremely small factor $ 10^{-53}$. Furthermore in this work, we would have to deal with the perturbations of this term. So, no doubt that if this term is of so small order, then its perturbations created due to the metric fluctuations generated from the charged particle or compact object, would be even smaller. So, it is now clear that the perturbative correction terms generated from this term $\frac{q^{2}}{3m}( R^{\mu }_{\lambda} u^{\lambda} + R^{\nu}_{\lambda} u^{\lambda} u^{\mu} u_{\nu} ) $ can be neglected. \\
On the other hand, the Abraham-Lorentz-Dirac term $ \frac{2 q^{2}}{3m} (\delta^{\mu}_{\nu} + u'^{\mu}u'_{\nu}) \frac{D^{2} u'^{\nu}}{d \tau^{2}} $ can not be expressed in the style of equation \ref{A2.7}, so the same can not be said for it. \\
For this reason, we are neglecting the perturbations generated from this term containing Ricci tensor in our work. 
% Furthermore, whether we are neglecting this term or not, that does not affect the analysis of the correction terms, which we have discussed in our work. 
%___________________________________________________________________________________________________________
\section{APPENDIX-3 : Certain explanations on comparing different parts within the correction terms, while investigating the significance of the correction terms : }
In this section, we clarify certain issues regarding the comparison of different parts of the corrections terms, given in equations \ref{4.1a} to \ref{4.5}, separately with the gravitational self-force term.
%In this section we explain why we donot need to compare all the parts of the correction terms, given in equations \ref{4.1a} to \ref{4.5}, separately with the gravitational self-force term .\\ 
%First of all, suppose we have to compare two quantities A and B, where A  = $A_{1} + A_{2} + A_{3} $. Then for $ A \sim B $ i.e. for $A_{1} + A_{2} + A_{3} \sim B $, one of the condition is $ A_{1} \sim B $. Now, if $A_{1} \sim B $ holds, even if $A_{2} $ and $A_{3} $ be very small or negligible in comparison with the quantity B, then also $A \sim  B$ will hold. This is true for any one among $A_{1}, \, A_{2} $ and $ A_{3} $. The only case, when despite of holding of  $A_{1} \sim B $, the $A \sim B $ would not hold, is when $A_{2} + A_{3} $ would be of similar order with $ A_{1}$, but with opposite sign. So, if the last-one does not holds, then $A_{1}  \sim B $ is a sufficient-condition for $A \sim B $. \\
%Therefore, in our work, if any one of terms in the numerators of the R.H.S. of equations \ref{4.6}, \ref{4.12} and \ref{5.1}, be of the order of that in the denominator i.e. the gravitational radiation reaction term, then the conditions will be satisfied. We just donot need to compare the other terms in the numerator. But, we do some of those too, to get the different conditions of their significance. \\
%Now, we explain why we have not compared all the terms in the numerators on the R.H.S. of the above mentioned  equations.
 In the numerator of the R.H.S. of the equation \ref{4.6}, it may seem that we have not compared the term $ \frac{2q^{2}}{m} \xi_{1}^{2} u'^{\alpha} \frac{d u'^{\beta}}{d \tau'} \Delta \Gamma^{\mu}_{\alpha \beta } $. But, actually if we compare both the terms together viz. $ \frac{2q^{2}}{m} \xi_{1}^{2} u'^{\alpha} \frac{d u'^{\beta}}{d \tau'} (\delta^{\mu}_{\eta} + u'^{\mu} u'_{\eta} ) \Delta \Gamma^{\eta}_{\alpha \beta } $ , in the numerator with that of the denominator viz. $ \xi_{1}^{2}  (\delta^{\mu}_{\eta} + u^{\mu} u_{\eta} )  $, then the resultant condition will be same as that, which is shown in the equation \ref{4.8}. [ It is to be noted that $ u'^{\mu} u'_{\eta}  \approx u^{\mu} u_{\eta}  $ ]. So, we do not need to separately compare the terms $ \frac{2q^{2}}{m} \xi_{1}^{2} u'^{\alpha} \frac{d u'^{\beta}}{d \tau'} u'^{\mu} u'_{\eta}  \Delta \Gamma^{\eta}_{\alpha \beta }  $ and $ \frac{2q^{2}}{m} \xi_{1}^{2} u'^{\alpha} \frac{d u'^{\beta}}{d \tau'}  \Delta \Gamma^{\mu}_{\alpha \beta } $ with that of the denominator. And, we have already stated that we have compared the term $ \frac{2q^{2}}{m} \xi_{1}^{2} u'^{\alpha} \frac{d u'^{\beta}}{d \tau'} h^{R \, \mu}_{\eta} \Gamma^{\eta}_{\alpha \beta} $, starting from equation \ref{4a}.  \\
 For the subsection 5.2 i.e. for comparison of the term $a^{\mu}_{4} $ with $a^{\mu}_{1} $, the similar argument is valid, as given above. If we compare the two terms : first and third ones together, in the numerator on the R.H.S. of the equation \ref{4.12}, with that of the denominator, then the ultimate result will be the same as that given in the equation \ref{4.19}. On the other hand, comparison of the rest term $\frac{2 q^{2}}{3 m } \xi_{1}^{2} u'^{\alpha} u'^{\beta} u'^{\gamma} h^{R \, \mu }_{\eta} $ with the denominator yields such a condition, that may not be predicted to be satisfied in any certain astrophysical or cosmological case. That is why, we have not compared that one.\\
 Similarly, in the subsection 5.3 i.e. for comparison of the term $a^{\mu}_{5} $ with $a^{\mu}_{1} $, same logic stands.
%___________________________________________________________________________________________________________

%___________________________________________________________________________________________________________
%___________________________________________________________________________________________________________
%___________________________________________________________________________________________________________
%___________________________________________________________________________________________________________
\end{small}
\end{document}